\documentclass{article}

\usepackage{xcolor}
\usepackage{amsmath}
\usepackage{graphicx}
\usepackage[inline,shortlabels]{enumitem}
\usepackage{newfloat}
\usepackage[skip=3pt]{subcaption}
\usepackage{xifthen}
\usepackage{hyperref}
\usepackage[capitalize]{cleveref}
\usepackage{quoting}
\usepackage{mdframed}
\usepackage{mathtools}
\usepackage{algorithm}
\usepackage[indLines=false]{algpseudocodex}
\usepackage{dirtree}
\usepackage{thmtools}
\usepackage{silence}
\newcommand{\techreport}{0}
\newcommand{\submission}{1}
\newcommand{\cameraready}{2}
\newcommand{\papermode}{\techreport}

\ifthenelse{\papermode = \techreport}{%
  \usepackage{amssymb}
  \usepackage[a4paper]{geometry}
  \usepackage[numbers]{natbib}
  \usepackage[T1]{fontenc}
  \usepackage[tt=false, type1=true]{libertine}
  \usepackage[varqu]{zi4}
  \usepackage[libertine]{newtxmath}
  \usepackage{varwidth}
  \usepackage[labelfont=bf]{caption}
  \usepackage{orcidlink}

  \usepackage[bottom]{footmisc}
}{}

\newcommand{\todo}[1]{\textcolor{red}{}}

\newcommand{\system}{\textsc{Basilisk}}

\newcommand{\pullup}{\textsc{TPullup}}
\newcommand{\pushdown}{\textsc{TPushdown}}
\newcommand{\iterpush}{\textsc{TIterPush}}
\newcommand{\pushand}{\textsc{TPushConj}}
\newcommand{\combined}{\textsc{TCombined}}
\newcommand{\tmin}{\textsc{TMin}}

\newcommand{\bdisj}{\textsc{BDisj}}
\newcommand{\bpushand}{\textsc{BPushConj}}

\newcommand{\propup}{\textsc{GeneralizeTag}}
\newcommand{\topmost}{\textsc{topmostAssignments}}
\newcommand{\calcbenefit}{\textsc{CalcBenefitScore}}
\newcommand{\canprop}{\textsc{canPropagate}}
\newcommand{\doprop}{\textsc{doPropagate}}

\newcommand{\inputtag}{\ensuremath{\left<in\mhyphen tag \right>}}
\newcommand{\postag}{\ensuremath{\left<pos\mhyphen tag \right>}}
\newcommand{\negtag}{\ensuremath{\left<neg\mhyphen tag \right>}}
\newcommand{\unktag}{\ensuremath{\left<unknown\mhyphen tag \right>}}
\newcommand{\lefttag}{\ensuremath{\left<left\mhyphen tag \right>}}
\newcommand{\righttag}{\ensuremath{\left<right\mhyphen tag \right>}}
\newcommand{\outtag}{\ensuremath{\left<out\mhyphen tag \right>}}

\newlength{\smallindent}
\newlength{\largeindent}
\newcommand{\phantombrace}{\phantom{\{}}

\newcommand{\ttag}[1]{\texttt{\small\textbf{\{#1\}}}}

\algnewcommand\True{\textbf{T}}
\algnewcommand\False{\textbf{F}}
\algnewcommand\Unknown{\textbf{U}}

\mathchardef\mhyphen="2D

\DeclareFloatingEnvironment{query}
\DeclareFloatingEnvironment{example}

\crefname{query}{Query}{Queries}
\crefname{example}{Example}{Examples}
\crefname{figure}{Figure}{Figures}

\setlist{leftmargin=*}

\DTsetlength{0.4em}{1em}{0.2em}{0.4pt}{0pt}

\newcommand{\relslice}[1][]{%
  \begingroup%
  \ifthenelse{\isin{c}{#1}}{%
    \def\x{Relational slice}%
  }{%
    \ifthenelse{\isin{C}{#1}}{%
      \def\x{Relational Slice}%
    }{%
      \def\x{relational slice}%
    }%
  }%
  \ifthenelse{\isin{s}{#1}}{%
    {\x}s%
    }{%
    {\x}%
  }%
  \endgroup%
}

\newcommand{\patom}[1][]{%
  \begingroup%
  \ifthenelse{\isin{c}{#1}}{%
    \def\x{Predicate}%
  }{%
    \ifthenelse{\isin{C}{#1}}{%
      \def\x{Predicate}%
    }{%
      \ifthenelse{\isin{a}{#1}}{%
        \def\x{P}%
      }{%
        \def\x{predicate}%
      }%
    }%
  }%
  \ifthenelse{\isin{s}{#1}}{%
    {\x}s%
    }{%
    {\x}%
  }%
  \endgroup%
}

\newcommand{\clause}[1][]{%
  \begingroup%
  \ifthenelse{\isin{c}{#1}}{%
    \def\x{Root clause}%
  }{%
    \ifthenelse{\isin{C}{#1}}{%
      \def\x{Root Clause}%
    }{%
      \def\x{root clause}%
    }%
  }%
  \ifthenelse{\isin{s}{#1}}{%
    {\x}s%
    }{%
    {\x}%
  }%
  \endgroup%
}

\ifthenelse{\papermode = \submission}{%
  \setcopyright{acmcopyright}
  \copyrightyear{2024}
  \acmYear{2024}
  \acmDOI{XXXXXXX.XXXXXXX}
  \acmConference[SIGMOD '24]{ACM SIGMOD}{June 09-15, 2024}{Santiago, Chile}
  \acmPrice{15.00}
  \acmISBN{978-1-4503-XXXX-X/18/06}
}{%
  \ifthenelse{\papermode = \cameraready}{%
    \setcopyright{rightsretained}
    \acmJournal{PACMMOD}
    \acmYear{2024} \acmVolume{2} \acmNumber{3 (SIGMOD)} \acmArticle{158} \acmMonth{6}\acmDOI{10.1145/3654961}
  }{}
}

\ifthenelse{\papermode = \techreport}{%
  \author{%
    Albert Kim\,\orcidlink{0009-0008-4692-0757}\\
    MIT \\
    {\tt alkim@csail.mit.edu} \and
    Samuel Madden\,\orcidlink{0000-0002-7470-3265} \\
    MIT \\
    {\tt madden@csail.mit.edu}
  }
  \date{}
}{%
  \author{Albert Kim}
  \email{alkim@csail.mit.edu}
  \orcid{0009-0008-4692-0757}
  \affiliation{%
    \institution{MIT}
    \city{Cambridge}
    \state{MA}
    \country{USA}
  }

  \author{Samuel Madden}
  \email{madden@csail.mit.edu}
  \orcid{0000-0002-7470-3265}
  \affiliation{%
    \institution{MIT}
    \city{Cambridge}
    \state{MA}
    \country{USA}
  }
}

\title{Optimizing Disjunctive Queries with Tagged Execution}

\begin{document}

%\begin{CCSXML}
%  <ccs2012>
%  <concept>
%  <concept_id>10002951.10002952.10003190.10003192</concept_id>
%  <concept_desc>Information systems~Database query processing</concept_desc>
%  <concept_significance>500</concept_significance>
%  </concept>
%  </ccs2012>
%\end{CCSXML}
%
%\ccsdesc[500]{Information systems~Database query processing}

\ifthenelse{\papermode = \techreport}{%
  \maketitle

  \begin{abstract}
  Despite  decades of research into query optimization, optimizing queries with disjunctive predicate expressions  remains a challenge.
  Solutions employed by existing systems (if any) are often simplistic and lead to much redundant work being performed by the execution engine.
  To address these problems, we propose a novel form of query execution called \emph{tagged execution}.
  Tagged execution groups tuples into subrelations based on which predicates in the query they satisfy (or don't satisfy) and tags them with that information.
  These tags then provide additional context for query operators to take advantage of during runtime, allowing them to eliminate much of the redundant work performed by traditional engines and realize predicate pushdown optimizations for disjunctive predicates.
  However, tagged execution brings its own challenges, and the question of what tags to create is a nontrivial one.
  Careless creation of tags can lead to an exponential blowup in the tag space, with the overhead outweighing the benefits.
  To address this issue, we present a technique called tag generalization to minimize the space of tags.
  We implemented the tagged execution model with tag generalization in our system \system{}, and our evaluation showed an average 2.7$\times$ speedup in runtime over the traditional execution model with up to a 19$\times$ speedup in certain situations.
\end{abstract}

}{%
  \begin{abstract}
  Despite  decades of research into query optimization, optimizing queries with disjunctive predicate expressions  remains a challenge.
  Solutions employed by existing systems (if any) are often simplistic and lead to much redundant work being performed by the execution engine.
  To address these problems, we propose a novel form of query execution called \emph{tagged execution}.
  Tagged execution groups tuples into subrelations based on which predicates in the query they satisfy (or don't satisfy) and tags them with that information.
  These tags then provide additional context for query operators to take advantage of during runtime, allowing them to eliminate much of the redundant work performed by traditional engines and realize predicate pushdown optimizations for disjunctive predicates.
  However, tagged execution brings its own challenges, and the question of what tags to create is a nontrivial one.
  Careless creation of tags can lead to an exponential blowup in the tag space, with the overhead outweighing the benefits.
  To address this issue, we present a technique called tag generalization to minimize the space of tags.
  We implemented the tagged execution model with tag generalization in our system \system{}, and our evaluation showed an average 2.7$\times$ speedup in runtime over the traditional execution model with up to a 19$\times$ speedup in certain situations.
\end{abstract}

  \ifthenelse{\papermode = \cameraready}{%

    \received{October 2023}
    \received[revised]{January 2024}
    \received[accepted]{February 2024}
  }{}

  \maketitle
}

\section{Introduction}
\label{sec:intro}

Despite decades of research in query optimization, optimizing queries with disjunctive predicates remains a relatively understudied problem.
Many systems and optimizers ignore disjunctions outright, and solutions (if any) employed by existing systems often fallback to simple and inefficient heuristics for evaluating disjunctive queries.
However, disjunctions continue to commonly appear in real workloads and pose issues for query optimization~\cite{fontouraEfficientlyEvaluatingComplex2010}~\cite{hennebergNorthlightDeclarativeOptimized2022}~\cite{albeAvoidBetterPostgreSQL2018}.

To illustrate the challenges, suppose we want to compile a list of potential movies to watch this weekend.
We are a fan of more recent movies because they have better special effects, so we are willing to watch them as long they have a score above 7.0.
However, we are also willing to tolerate the effects of older movies if they are ``masterpieces'' and have a score greater than 8.0.
\cref{query:ex}\ifthenelse{\papermode = \cameraready}{ (next page)}{} expresses these constraints.
Note the schema for this query comes from the IMDB dataset provided by the Join Order Benchmark~\cite{leisHowGoodAre2015}, and the query is a simplification of one of the queries in the benchmark\footnote{We take a few liberties with the attribute names and predicate expressions in this example to improve brevity and clarity.}.

\begin{query}
  \ifthenelse{\papermode = \techreport}{}{\vspace{.2in}}
  \ttfamily
  \centering
  \begin{varwidth}{1.1\linewidth}
  \begin{tabbing}
    SELECT * FROM \= title AS t JOIN movie\_info\_idx AS mi\_idx \\
    \> ON t.id = mi\_idx.movie\_id \\
    \> WHERE\' (t.year > 2000 AND mi\_idx.score > '7.0') \\
    \> OR\' (t.year > 1980 AND mi\_idx.score > '8.0')
  \end{tabbing}
  \end{varwidth}
  %\caption{Example query in style of join order benchmark.}
  \ifthenelse{\papermode = \submission}{\vspace{-.1in}}{}
  \caption{}
  \label{query:ex}
  \ifthenelse{\papermode = \techreport}{}{\vspace{-.2in}}
\end{query}

Due to the disjunction in \cref{query:ex}'s predicate expression, performing pushdown optimizations is not straightforward.
In this case, existing systems typically do one of two things:
\begin{enumerate*}
  \item Perform the join first, then evaluate the predicate expression on the resulting joined output.
  \item Treat each part of the disjunction as a separate query, applying pushdown optimizations separately, and combine the results with a union operator.
    Note this is equivalent to internally transforming \cref{query:ex} into:
\end{enumerate*}
\smallskip
\begin{quoting}[leftmargin=0pt]
  \ifthenelse{\papermode = \submission}{\vspace{-.5em}}{}
  \ttfamily
  \centering
  \begin{varwidth}{1.1\linewidth}
  \begin{tabbing}
    SELECT * FROM \= title AS t JOIN movie\_info\_idx AS mi\_idx \\
    \> ON t.id = mi\_idx.movie\_id \\
    \> WHERE\' t.year > 2000 AND mi\_idx.score > '7.0' \\
    UNION \\
    SELECT * FROM title AS t JOIN movie\_info\_idx AS mi\_idx \\
    \> ON t.id = mi\_idx.movie\_id \\
    \> WHERE\' t.year > 1980 AND mi\_idx.score > '8.0'
  \end{tabbing}
  \end{varwidth}
  \ifthenelse{\papermode = \submission}{\vspace{.1em}}{}
\end{quoting}
\smallskip

Both solutions are unsatisfactory.
The first option is equivalent to performing no optimizations whatsoever.
\cref{query:ex} only features a single join, but without a way to prune the relations before the join, as the number of joins grows, the size of the joined output will grow exponentially and the total runtime alongside with it.
The second option allows for pushdown optimizations, but does redundant work because movies that scored above 8.0 and were made after 2000 satisfy both parts of the disjunction.
The tuples representing these movies need to be constructed multiple times.
An additional, potentially expensive union operator is also required to filter out duplicate tuples.
It is also worth noting that the second option is only even available because \cref{query:ex}'s predicate expression is a disjunct of conjuncts (DNF).
If it were a conjunct of disjuncts (CNF), the second option would not be available.

To address these problems, we propose a novel form of query execution called \emph{tagged execution}.
In traditional query execution, operators such as filters and joins operate on and result in sets of tuples called relations.
In tagged execution, tags loaded with semantic information are attached to subsets of relations, and operators use that extra information to avoid performing any redundant work during query execution, thereby significantly improving runtime performance.
The tagged execution model allows the query optimizer to push down disjunctive predicates regardless of whatever form the predicate expression might have.
In addition, it ensures that each tuple is only ever materialized once and that each predicate subexpression is only evaluated once, even if it appears multiple times in the overall predicate expression.

Optimally executing \cref{query:ex} under tagged execution would:
\begin{enumerate}[nosep]
  \item Apply predicates \texttt{t.year > 2000} and \texttt{t.year > 1980} to the \texttt{title} table and attach the tags \ttag{t.year > 2000 = T} and \ttag{t.year > 2000 = F, t.year > 1980 = T} to the relevant tuples.
  \item Apply predicates \texttt{mi\_idx.score > '8.0'} and \texttt{mi\_idx.score > '7.0'} to the \texttt{movie\_info\_idx} table and attach the tags \ttag{mi\_idx.score > '8.0' = T} and \ttag{mi\_idx.score > '8.0' = F, mi\_idx.score > '7.0' = T} to the relevant tuples.
  \item Do the following joins between the sets of tuples with tags:
    \begin{enumerate}
      \item \ttag{t.year > 2000 = T} $\times$ \ttag{mi\_idx.score > '8.0' = T}
      \item \ifthenelse{\papermode = \submission}{%
          \ttag{t.year > 2000 = T} $\times$ \\ \strut \ttag{mi\_idx.score > '8.0' = F, mi\_idx.score > '7.0' = T}
        }{%
          \ttag{t.year > 2000 = T} $\times$ \ttag{mi\_idx.score > '8.0' = F, mi\_idx.score > '7.0' = T}
        }
      \item \ifthenelse{\papermode = \submission}{%
          \ttag{t.year > 2000 = F, t.year > 1980 = T} $\times$ \\ \strut \ttag{mi\_idx.score > '8.0' = T}
        }{%
          \ttag{t.year > 2000 = F, t.year > 1980 = T} $\times$ \ttag{mi\_idx.score > '8.0' = T}
        }
    \end{enumerate}
\end{enumerate}
Note the predicates are pushed down, and the join only ever processes each tuple once.

\noindent
\textbf{Challenges.}
Although tagged execution is a powerful new paradigm for query execution, it introduces new technical challenges:
\begin{enumerate}[nosep]
  \item
    \textbf{Tag Management.}
    How should tags be generated, which should be preserved, and how should they be combined?
    A naive implementation storing all true/false assignment values of predicates can lead to an exponential number of tags, so we must manage tags carefully to ensure overheads do not outweigh the benefits.
  \item
    \textbf{Planning.}
    A new execution model requires a new query planner.
    We must explore into how much of conventional planning wisdom we can bring into tagged execution and devise new planners which take advantage of the unique benefits offered by tagged execution.
\end{enumerate}

\noindent
\textbf{Contributions.}
In short our contributions are:
\begin{enumerate}[nosep]
  \item
    The tagged execution model, with its ability to optimize and push down disjunctive predicates.
  \item
    Our solution for the tag management problem and several new planners to use with tagged execution.
  \item
    The evaluation of tagged execution in our system \system{}\footnote{The common basilisk can run across water by ``pushing down'' with its feet rapidly enough to create air pockets to push off of~\cite{hsiehRunningWaterThreedimensional2004}!}, in which tagged execution achieved an average 2.7$\times$ speedup in runtime over traditional query execution with up to a 19$\times$ speedup in certain situations.
\end{enumerate}
The rest of this paper is structured as follows.
\cref{sec:model} provides a detailed explanation of the tagged execution model.
%\cref{sec:tag-space} introduces a technique called tag generalization to serve as our solution to the tag management problem.
\cref{sec:tag-space} introduces a technique called tag generalization to solve the tag management problem and discusses how to handle NULL values.
\cref{sec:plan} provides an overview into planning for tagged execution.
\cref{sec:eval} presents our evaluation, \cref{sec:rel-work} discusses related work, and \cref{sec:conc} concludes the paper.

\todo{remove balance warning filter}

\section{Tagged Execution}
\label{sec:model}

In this section, we describe the tagged execution model.
We introduce the concept of tags and describe how operators in tagged execution use these tags to avoid redundant work.
It should be noted that the actual creation of tags and the decision of which tags to create are all done and made by the planner during plan time, which we describe in \cref{sec:tag-space}.
This section only focuses on how the execution engine performs tagged execution during runtime, given a query plan and some set of tags.

\begin{figure}
  \centering
  \ifthenelse{\papermode = \submission}{%
    \includegraphics[width=.8\linewidth]{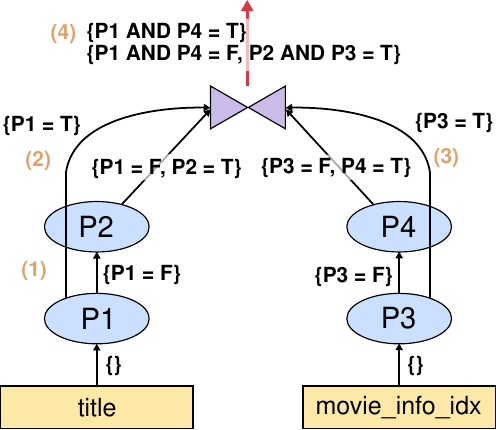}
  }{%
    \includegraphics[width=.6\linewidth]{figs/query-plan2.pdf}
  }
  \caption{Example tagged execution query plan of \cref{query:ex}. Here, P1 = \texttt{t.year > 2000}, P2 = \texttt{t.year > 1980}, P3 = \texttt{mi\_idx.score > '8.0'}, and P4 = \texttt{mi\_idx.score > '7.0'}.}
  \label{fig:query-plan}
  \vspace{-10pt}
\end{figure}

As an overarching example for this section, consider \cref{fig:query-plan}\ifthenelse{\papermode = \cameraready}{ (next page)}{}, which shows a sample query plan for the tagged execution of \cref{query:ex}.
The base table nodes \texttt{title} and \texttt{movie\_info\_idx} feed into the four base predicates from \cref{query:ex} before coming together for a final join.  As we describe next, this plan successfully achieves disjunctive predicate pushdown.
The orange numbers in parentheses refer to \cref{ex:title1,ex:title2,ex:mi-idx,ex:joined} \ifthenelse{\papermode = \cameraready}{(\cpageref{ex:title1})}{(next page)}, which list example tuples for each of these stages.
Focusing on the left side, the tuples from \texttt{title} are originally assigned the empty tag (i.e., \ttag{}).
After applying the predicate \texttt{t.year > 2000}, the tuples which evaluate to true are assigned the tag \ttag{t.year > 2000 = T}, and the tuples which evaluate to false are assigned the tag \ttag{t.year > 2000 = F}.
The second filter can then avoid redundant work by applying the predicate \texttt{t.year > 1980} on only the tuples with the tag \ttag{t.year > 2000 = F}; the tuples with the tag \ttag{t.year > 2000 = T} can pass through the filter untouched.
Note the second filter does not output the tuples which would be assigned the tag \ttag{t.year > 2000 = F, t.year > 1980 = F}.
These tuples do not satisfy the overall predicate expression, so they can be dropped here and do not need to be fed into the join.
After performing similar actions on the right side, the join operator then uses the tags to selectively join only the tuples which satisfy the overall predicate expression.
In particular, the tuples which would be associated with the tag \ttag{t.year > 2000 = F, t.year > 1980 = T, mi\_idx.score > '8.0' = F, mi\_idx.score > '7.0' = T} are never joined. %, since they do not satisfy the overall predicate expression.
Thus, by relying on tags to keep track of predicate results, tagged execution is able to avoid much of the redundant work that would be performed by the traditional execution model.

In the remainder of this section, we first define the concept of relations with tags in \cref{sec:model:setup} and then describe how operators uses these tags in tagged execution in \cref{sec:model:filter,sec:model:join,sec:model:proj}.
We wrap up with some important implementation details regarding tagged execution in \cref{sec:model:impl}.

%SPACE - above paragraph

\subsection{Setup}
\label{sec:model:setup}
In traditional query execution, operators process and produce sets of tuples, or \emph{relations}.
In tagged execution, this basic unit of operation becomes \emph{tagged relations}.
Tagged relations are similar to regular relations in that they comprise of sets of tuples, but subsets of tuples in tagged relations (called \emph{\relslice[s]}) are annotated with tags containing semantic information, providing additional context for operators to take advantage of during query execution.
%Alternatively, a tagged relation is essentially a set of multiple ``mini-relations'' which operators can pass to each other to avoid performing redundant work.
A tagged relation may have any number of \relslice[s].
However, the \relslice[s] must be mutually exclusive, and every \relslice[] must be associated with exactly one tag.
Although the model allows for \relslice[s] with zero tuples, in practice, these \relslice[s] are removed from tagged relations for performance reasons.
%In other words, each tuple in a tagged relation is associated with exactly one tag; tuples not associated with any tags are removed from the tagged relation.
The tags themselves are sets of true/false \emph{assignments} to arbitrarily complex predicate expressions from the query, and an assignment has the form:
\[
  \langle expr \rangle = T/F
\]
In which $\langle expr \rangle$ is an arbitrarily complex Boolean SQL expression, and an assignment to $T$ represents $\langle expr \rangle$ is true, and an assignment to $F$ represents $\langle expr \rangle$ is false.
Each tag may have any number of assignments, and each tuple in the corresponding  \relslice[] must satisfy every assignment present in the associated tag.
For example, every tuple in the \relslice[] associated with the tag \ttag{t.year > 2000 = F, t.year > 1980 = T} must be a title produced between 1981 and 2000.
%\srm{Sets of assignments to conjunctive application of assignments is confusing -- an example would help I think}
%(a ``nonce'' tag which contains no assignments is allowed). \srm{Not sure what ``nonce'' means here or if we need to say this - it's a bit confusing.}
%As we show later in \cref{sec:tag-space}, assignments to complex predicate expressions is precisely what allows us to reduce the tag space.

%Since tags refer to predicate expressions, tagged relations are often specific to a certain query and its predicate expression.  \srm{I feel this is perhaps trying to be arbitrarily general -- can we just describe in terms of boolean expressions over fields?}

\cref{ex:title1,ex:title2,ex:mi-idx,ex:joined} are all examples of tagged relations associated with \cref{query:ex}.
The schema for the tuples are shown in the captions.
Tags are in bold, and corresponding \relslice[s] are given in the following non-bold lines.
%True/false assignments to predicate expressions in tags are depicted with a \texttt{\textbf{= T}/\textbf{= F}} after the predicate expression.
Looking at \cref{ex:title1}, we see that the tagged relation is made up of 7 tuples split into 2 \relslice[s]; the two \relslice[s] divide the tuples based on whether they were produced after 2000 or not.
%\cref{ex:title2,ex:mi-idx,ex:joined} all depict multiples assignments within a single tag, and \cref{ex:joined} features an assignment to a complex predicate expression (including an AND).
\cref{ex:joined} features an example of an assignment to a \emph{complex} predicate expression, which we define as a predicate expression containing either an AND or an OR.
Note that in all examples, no tuple satisfies more than a single set of assignments present in a tagged relation, satisfying mutual exclusivity.

\ifthenelse{\papermode = \submission}{%
\begin{example}
  \small
  \ttfamily
  \flushleft
  \ttag{t.year > 2000 = T}: \\
  \hspace*{\smallindent} \{('The Dark Knight', 2008, 1), ('Evolution', 2001, 2), \\
  \hspace*{\smallindent} \phantombrace{}('Avatar', 2009, 7)\} \\
  \strut \ttag{t.year > 2000 = F}: \\
  \hspace*{\smallindent} \{('The Shawshank Redemption', 1994, 3),\\
  \hspace*{\smallindent} \phantombrace{}('Pulp Fiction', 1994, 4), ('The Godfather', 1972, 5)\\
  \hspace*{\smallindent} \phantombrace{}('Beetlejuice', 1988, 6)\}
  \caption{Pre-filter \texttt{title(title, year, id)}}
  \label{ex:title1}
  \vspace{-4pt}
\end{example}

\begin{example}
  \small
  \ttfamily
  \flushleft
  \ttag{t.year > 2000 = T}: \\
  \hspace*{\smallindent} \{('The Dark Knight', 2008, 1), ('Evolution', 2001, 2), \\
  \hspace*{\smallindent} \phantombrace{}('Avatar', 2009, 7)\} \\
  \strut \ttag{t.year > 2000 = F, t.year > 1980 = T}: \\
  \hspace*{\smallindent} \{('The Shawshank Redemption', 1994, 3),\\
  \hspace*{\smallindent} \phantombrace{}('Pulp Fiction', 1994, 4), ('Beetlejuice', 1988, 6)\} \\
  \caption{Post-filter \texttt{title(title, year, id)}}
  \label{ex:title2}
  \vspace{-4pt}
\end{example}

\begin{example}
  \small
  \ttfamily
  \flushleft
  \ttag{mi\_idx.score > '8.0' = T} \\
  \hspace*{\smallindent} \{('9.0', 1),  ('9.3', 3), ('8.9', 4), ('9.2', 5)\} \\
  \strut \ttag{mi\_idx.score > '8.0' = F, mi\_idx.score > '7.0' = T} \\
  \hspace*{\smallindent} \{('7.5', 6),  ('7.9', 7)\} \\
  %\strut \ttag{mi\_idx.score > '8.0' = F, mi\_idx.score > '7.0' = F} \\
  %\hspace*{\smallindent} \{('6.1', 2)\}
  \caption{\texttt{movie\_info\_idx(score, movie\_id)}}
  \label{ex:mi-idx}
  \vspace{-4pt}
\end{example}

\begin{example}
  \small
  \ttfamily
  \flushleft
  \ttag{t.year > 2000 AND mi\_idx.score > '7.0' = T}: \\
  \hspace*{\smallindent} \{('The Dark Knight', 2008, '9.0', 1), \\
  \hspace*{\smallindent} \phantombrace{}('Avatar', 2009, '7.9', 7)\} \\
  \strut \ttag{t.year > 2000 AND mi\_idx.score > '7.0' = F, \\ \phantombrace{}t.year > 1980 AND mi\_idx.score > '8.0' = T}: \\
  \hspace*{\smallindent} \{('The Shawshank Redemption', 1994, '9.3', 3),\\
  \hspace*{\smallindent} \phantombrace{}('Pulp Fiction', 1994, '8.9', 4)\}
  \caption{Joined \texttt{(title, year, score, id)}}
  \label{ex:joined}
  \vspace{-12pt}
\end{example}
}{%
  \ifthenelse{\papermode = \techreport}{%
    \begin{example}[H]
      \small
      \ttfamily
      \flushleft
      \ttag{t.year > 2000 = T}: \\
      \hspace*{\smallindent} \{('The Dark Knight', 2008, 1), ('Evolution', 2001, 2), ('Avatar', 2009, 7)\} \\
      \strut \ttag{t.year > 2000 = F}: \\
      \hspace*{\smallindent} \{('The Shawshank Redemption', 1994, 3), ('Pulp Fiction', 1994, 4), \\
      \hspace*{\smallindent} \phantombrace{}('The Godfather', 1972, 5), ('Beetlejuice', 1988, 6)\}
      \caption{Pre-filter \texttt{title(title, year, id)}}
      \label{ex:title1}
      \vspace{-4pt}
    \end{example}

    \begin{example}[H]
      \small
      \ttfamily
      \flushleft
      \ttag{t.year > 2000 = T}: \\
      \hspace*{\smallindent} \{('The Dark Knight', 2008, 1), ('Evolution', 2001, 2), ('Avatar', 2009, 7)\} \\
      \strut \ttag{t.year > 2000 = F, t.year > 1980 = T}: \\
      \hspace*{\smallindent} \{('The Shawshank Redemption', 1994, 3), ('Pulp Fiction', 1994, 4), \\
      \hspace*{\smallindent} \phantombrace{} ('Beetlejuice', 1988, 6)\}
      \caption{Post-filter \texttt{title(title, year, id)}}
      \label{ex:title2}
      \vspace{-4pt}
    \end{example}

    \begin{example}[H]
      \small
      \ttfamily
      \flushleft
      \ttag{mi\_idx.score > '8.0' = T} \\
      \hspace*{\smallindent} \{('9.0', 1),  ('9.3', 3), ('8.9', 4), ('9.2', 5)\} \\
      \strut \ttag{mi\_idx.score > '8.0' = F, mi\_idx.score > '7.0' = T} \\
      \hspace*{\smallindent} \{('7.5', 6),  ('7.9', 7)\} \\
      %\strut \ttag{mi\_idx.score > '8.0' = F, mi\_idx.score > '7.0' = F} \\
      %\hspace*{\smallindent} \{('6.1', 2)\}
      \caption{\texttt{movie\_info\_idx(score, movie\_id)}}
      \label{ex:mi-idx}
      \vspace{-4pt}
    \end{example}

    \begin{example}[H]
      \small
      \ttfamily
      \flushleft
      \ttag{t.year > 2000 AND mi\_idx.score > '7.0' = T}: \\
      \hspace*{\smallindent} \{('The Dark Knight', 2008, '9.0', 1), ('Avatar', 2009, '7.9', 7)\} \\
      \strut \ttag{t.year > 2000 AND mi\_idx.score > '7.0' = F, t.year > 1980 AND mi\_idx.score > '8.0' = T}: \\
      \hspace*{\smallindent} \{('The Shawshank Redemption', 1994, '9.3', 3), ('Pulp Fiction', 1994, '8.9', 4)\}
      \caption{Joined \texttt{(title, year, score, id)}}
      \label{ex:joined}
      \vspace{-12pt}
    \end{example}
  }{%
    \begin{example}
      \small
      \ttfamily
      \flushleft
      \ttag{t.year > 2000 = T}: \\
      \hspace*{\smallindent} \{('The Dark Knight', 2008, 1), ('Evolution', 2001, 2), ('Avatar', 2009, 7)\} \\
      \strut \ttag{t.year > 2000 = F}: \\
      \hspace*{\smallindent} \{('The Shawshank Redemption', 1994, 3), ('Pulp Fiction', 1994, 4), \\
      \hspace*{\smallindent} \phantombrace{}('The Godfather', 1972, 5), ('Beetlejuice', 1988, 6)\}
      \caption{Pre-filter \texttt{title(title, year, id)}}
      \label{ex:title1}
      \vspace{-4pt}
    \end{example}

    \begin{example}
      \small
      \ttfamily
      \flushleft
      \ttag{t.year > 2000 = T}: \\
      \hspace*{\smallindent} \{('The Dark Knight', 2008, 1), ('Evolution', 2001, 2), ('Avatar', 2009, 7)\} \\
      \strut \ttag{t.year > 2000 = F, t.year > 1980 = T}: \\
      \hspace*{\smallindent} \{('The Shawshank Redemption', 1994, 3), ('Pulp Fiction', 1994, 4), \\
      \hspace*{\smallindent} \phantombrace{} ('Beetlejuice', 1988, 6)\}
      \caption{Post-filter \texttt{title(title, year, id)}}
      \label{ex:title2}
      \vspace{-4pt}
    \end{example}

    \begin{example}
      \small
      \ttfamily
      \flushleft
      \ttag{mi\_idx.score > '8.0' = T} \\
      \hspace*{\smallindent} \{('9.0', 1),  ('9.3', 3), ('8.9', 4), ('9.2', 5)\} \\
      \strut \ttag{mi\_idx.score > '8.0' = F, mi\_idx.score > '7.0' = T} \\
      \hspace*{\smallindent} \{('7.5', 6),  ('7.9', 7)\} \\
      %\strut \ttag{mi\_idx.score > '8.0' = F, mi\_idx.score > '7.0' = F} \\
      %\hspace*{\smallindent} \{('6.1', 2)\}
      \caption{\texttt{movie\_info\_idx(score, movie\_id)}}
      \label{ex:mi-idx}
      \vspace{-4pt}
    \end{example}

    \begin{example}
      \small
      \ttfamily
      \flushleft
      \ttag{t.year > 2000 AND mi\_idx.score > '7.0' = T}: \\
      \hspace*{\smallindent} \{('The Dark Knight', 2008, '9.0', 1), ('Avatar', 2009, '7.9', 7)\} \\
      \strut \ttag{t.year > 2000 AND mi\_idx.score > '7.0' = F, t.year > 1980 AND mi\_idx.score > '8.0' = T}: \\
      \hspace*{\smallindent} \{('The Shawshank Redemption', 1994, '9.3', 3), ('Pulp Fiction', 1994, '8.9', 4)\}
      \caption{Joined \texttt{(title, year, score, id)}}
      \label{ex:joined}
      \vspace{-12pt}
    \end{example}
}}

\subsection{Filter}
\label{sec:model:filter}
Given the previous setup, we can now describe how filter operators function in tagged execution.
In traditional query execution, a filter operator applies a predicate expression to an input relation and outputs the subset of tuples which evaluate to true for that predicate expression.
In tagged execution, each filter operator is given a \emph{tag map} specifying  which \relslice[s] of the input tagged relation it should evaluate the predicate expression on and what to do with the results.
An entry in the tag map has the following signature:
\[
  \inputtag \rightarrow \{T: Optional \postag,\; F: Optional \negtag\}
\]
The predicate expression is applied to all \relslice[s] which have a tag that matches \inputtag{}.
If the optional \postag{} is specified, then the tuples which evaluate to true are stored as a \relslice[] in the output tagged relation with the tag \postag{}.
Similarly, if \negtag{} is specified, the tuples which evaluate to false form a \relslice[] with the tag \negtag{}.
Both \postag{} and \negtag{} may be specified in a single entry.
All \relslice[s] which do not have any matching entries in the tag map are passed untouched to the output tagged relation.
Note that same tag may appear as an output in multiple entries.
In this case, \relslice[s] which share the same output tag are merged together in the output tagged relation.

As an example, consider the filter operator with the predicate expression \texttt{t.year > 1980} being given a tag map with one entry:
\smallskip
\begin{quoting}[leftmargin=0em]
  \ifthenelse{\papermode = \submission}{%
    \ttag{t.year > 2000 = F} \\
    \hspace*{\largeindent} $\rightarrow \{T:$ \ttag{t.year > 2000 = F, t.year > 1980 = T} \}
  }{%
    \centering
    \ttag{t.year > 2000 = F} $\rightarrow \{T:$ \ttag{t.year > 2000 = F, t.year > 1980 = T} \}
  }
\end{quoting}
\smallskip
When the tagged relation from \cref{ex:title1} is passed in as input, only the second \relslice[] has a matching entry.
Thus, the predicate expression is only evaluated on the tuples from that \relslice[].
%The tuples which evaluate to true (i.e., ``The Shawshank Redemption'', ``Pulp Fiction'', and ``Beetlejuice'') come together to form a new \relslice[] with the tag \ttag{t.year > 2000 = F, t.year > 1980 = T}.
The tuples which evaluate to true come together to form a new \relslice[] with the tag \ttag{t.year > 2000 = F, t.year > 1980 = T}.
%Meanwhile, the \negtag{} is missing from the tag entry, so tuples which evaluate to false (i.e., ``The Godfather'') are removed and not included in the output tagged relation.
Meanwhile, the \negtag{} is missing from the tag entry, so tuples which evaluate to false are removed and not included in the output tagged relation.
The first \relslice[] does not match any entries in the tag map and is passed untouched to the output tagged relation.
The resulting tagged relation is shown in \cref{ex:title2}.
%Note that all movies which were produced after 2000 were also produced after 1980, so it would be meaningless to evaluate the predicate expression on the first \relslice[].
%This tag map is an example of how a sophisticated planner can reduce the work done by limiting the \relslice[s] which the predicate expression is evaluated on.

The planner determines the tag map for each filter operator.
During runtime, the execution engine simply follows the instructions encoded in the tag map.
As such, the degree of sophistication of the planner can have a large impact on performance, and a naive planner, such as one that produces a tag map entry for every \relslice[] in the input tagged relation and outputs all resulting true/false \relslice[s], can lead to  poor runtimes.
Thus, we show in \cref{sec:tag-space} how our planners build effective tag maps for efficient execution.
In the case of this example, the planner was intelligent enough to realize that titles produced after 2000 are also produced after 1980 and reduces the filter operator's work by omitting the tag map entry for \ttag{t.year > 2000 = T}.

%While this would enable the pushdown of disjunctive predicates, it would also be rather inefficient.
%The number of tags needed would be exponential in the number of filter operators, and predicate expressions would be evaluated on \relslice[s] for which the result would be meaningless.
%A more sophisticated planner would take care to minimize the tag space while ensuring predicate expressions are only evaluated on \relslice[s] which need them.
%\srm{Give forward reference to planner section.}
%Note that all movies which were produced after 2000 were also produced after 1980, so it would be meaningless to evaluate the predicate expression on the first \relslice[].
%This tag map is an example of how a sophisticated planner can reduce the work done by limiting the \relslice[s] which the predicate expression is evaluated on.

\subsection{Join}
\label{sec:model:join}
Similar to filter operators, join operators in tagged execution are given tag maps which determine how to combine their inputs during runtime.
Assuming a join operator has an input ``left'' tagged relation and an input ``right'' tagged relation, an entry in the tag map has the following signature:
\[
  (\lefttag, \righttag) \rightarrow \outtag
\]
For every pairing between a left \relslice[] and a right \relslice[], if it has a matching $(\lefttag,\allowbreak \righttag)$ entry, the tuples in those \relslice[s] are joined to create an output \relslice[] with the tag \outtag{}.
Similar to filters, output \relslice[s] which share the same tag are merged together in the output tagged relation.
However, unlike filter operators, \relslice[s] without a matching tag map entry are discarded and not included in the output tagged relation.
Note the same join constraint is used for every pairing of \relslice[s].

As an example, consider the join between the tagged relations in \cref{ex:title2} (left) and \cref{ex:mi-idx} (right) using the join constraint \texttt{title.id = movie\_info\_idx.movie\_id} given a tag map with the following entries:
\smallskip
\begin{quoting}[leftmargin=0em]
  \ifthenelse{\papermode = \submission}{%
    (\ttag{t.year > 2000 = T}, \ttag{mi\_idx.score > '8.0' = T}) \\
    \hspace*{\largeindent} $\rightarrow$ \ttag{t.year > 2000 AND mi\_idx.score > '7.0' = T} \\
    (\ttag{t.year > 2000 = T}, \\
    \hspace*{\smallindent} \ttag{mi\_idx.score > '8.0' = F, mi\_idx.score > '7.0' = T}) \\
    \hspace*{\largeindent} $\rightarrow$ \ttag{t.year > 2000 AND mi\_idx.score > '7.0' = T} \\
    (\ttag{t.year > 2000 = F, t.year > 1980 = T}, \\
    \hspace*{\smallindent} \ttag{mi\_idx.score > '8.0' = T}) \\
    \hspace*{\largeindent} $\rightarrow$ \ttag{t.year > 2000 AND mi\_idx.score > '7.0' = F,\\* \hspace*{\largeindent} \phantom{$\rightarrow$} t.year > 1980 AND mi\_idx.score > '8.0' = T}
  }{%
    \centering
    \begin{varwidth}{1.1\linewidth}
    (\ttag{t.year > 2000 = T}, \ttag{mi\_idx.score > '8.0' = T}) \\
    \hspace*{\largeindent} $\rightarrow$ \ttag{t.year > 2000 AND mi\_idx.score > '7.0' = T} \\
    (\ttag{t.year > 2000 = T},  \ttag{mi\_idx.score > '8.0' = F, mi\_idx.score > '7.0' = T}) \\
    \hspace*{\largeindent} $\rightarrow$ \ttag{t.year > 2000 AND mi\_idx.score > '7.0' = T} \\
    (\ttag{t.year > 2000 = F, t.year > 1980 = T},  \ttag{mi\_idx.score > '8.0' = T}) \\
    \hspace*{\largeindent} $\rightarrow$ \ttag{t.year > 2000 AND mi\_idx.score > '7.0' = F,\\* \hspace*{\largeindent} \phantom{$\rightarrow$} t.year > 1980 AND mi\_idx.score > '8.0' = T}
  \end{varwidth}
  }
\end{quoting}
\smallskip
The first and second entries join the first \relslice[] from \cref{ex:title2} with the first and second \relslice[s] respectively from \cref{ex:mi-idx} to generate the \relslice[] with tag \ttag{t.year > 2000 AND mi\_idx.score > '7.0' = T}.
The third entry joins the second \relslice[] from \cref{ex:title2} with the first \relslice[] from \cref{ex:mi-idx} to produce the \relslice[] with movies produced after 1980 with a score above 8.0 not included in the first output \relslice[].
Just as with filter operators, the planner determines the tag map, and intelligent planners can choose to selectively join only the tuples that satisfy the overall predicate expression and avoid performing costly joins between \relslice[s] whose results would be thrown away later in the pipeline.
In this example, the entry (\ttag{t.year > 2000 = F, t.year > 1980 = T}, \ttag{mi\_idx.score > '8.0' = F, mi\_idx.score > '7.0' = T}) is omitted from the tag map because the planner recognizes that movies produced between 1981 and 2000 with a score between 7.1 and 8.0 do not satisfy the overall predicate expression.
%In this case, movies produced between 1981 and 2000 with a score between 7.1 and 8.0 do not satisfy the overall predicate expression, and the planner can avoid unnecessary work by omitting the aforementioned tag map entry.

\subsection{Projection}
\label{sec:model:proj}
A projection operator in tagged execution offers one final chance to filter based on tags before actual projection.
The operator is given a set of tags by the planner, and only the tuples in \relslice[s] which have a matching tag are projected.
In the case of \cref{ex:joined}, all \relslice[s] are required as results to \cref{query:ex}, so the planner would include both tags for the projection operator.

\subsection{Implementation}
\label{sec:model:impl}
The exact implementation of tagged relations and the previously discussed operators can have a large impact on the overall runtime performance.
Here, we discuss the details of how tagged execution is implemented in our system \system{}.

\subsubsection{Tagged Relation}
\label{sec:model:impl:tag-rel}
\system{} is a column-oriented system, so intermediate representations of relations contain tuples of indices rather than tuples of actual values.
Each $n$-tuple is the result of joining $n$ tuples and contains indices into the $n$ tables.
The actual tuple of values can be reconstructed by doing an index-based lookup into each table when needed.
%Partial reconstruction is also possible if only specific attributes from certain tables are needed.
Given a relation of such tuples, tagged relations are constructed by creating an accompanying hash table of bitmaps.
Tags serve as keys to the hash table, and each bitmap specifies which tuples belong to which \relslice[].
Although an alternative implementation of separating out each \relslice[] into its own relation was considered, we found this version to be less performant because it required moving around tuples as opposed to manipulating bits in a bitmap.
% Separating out \relslice[s] often put indices out of order when reading, and this took extra time

\subsubsection{Filter}
There are two important implementation details for filter operators.
First, when evaluating the predicate expression, \system{} takes the union across all bitmaps which have matching tags, and the predicate expression is evaluated once on the tuples specified by the combined bitmap.
This results in fewer I/O calls to read the underlying data values than evaluating the predicate expression separately for each \relslice[].
Second, filter operators do not actually modify the underlying, non-tagged relation. %; the input relation becomes the output relation as-is.
Instead, only the hash table of bitmaps is updated to mirror the output tags and corresponding \relslice[s].
Even tuples which no longer belong to any \relslice[] remain in the relation because removing a tuple from the middle of a relation would require expensive modifications to every single bitmap.

\subsubsection{Join}
\system{} uses a hash join for all its joins.
However, rather than building a separate hash table for each \relslice[], \system{} builds one giant hash table for all the \relslice[s].
The values of the hash table contain enough supplementary details to determine which \relslice[s] the key belongs to.
This significantly improves runtime performance because \relslice[s] can share join key values and only one hash table needs to be allocated.

%\todo{Maybe make clear tagged execution supersedes traditional execution, and we can make the tagged operators do exactly what traditional operators would do}

\section{Tag Management}
\label{sec:tag-space}

The previous section described the mechanism by which tagged execution query operators can use tags during runtime to process only a subset of the input tagged relations.
We now turn our attention to the problem of tag management; for each query, how should the planner decide which tags to use, and how should it build each tag map.
This is important, since the tag maps ultimately dictate how much work is done for each query, and a naive strategy to tag management can lead to an exponential number of tags, causing the overhead of tagged execution to outweigh the benefits.
To demonstrate, we first describe this naive strategy in \cref{sec:tag-space:naive}, and we show that while it is capable of accomplishing disjunctive predicate pushdown, its weaknesses make it inferior to traditional execution strategies.
We then introduce the concept of \emph{generalizing} tags in \cref{sec:tag-space:general} and show that by using these generalized tags instead of the tags in the naive strategy, we can reduce the total number of tags in the system and avoid the exponential blowup in many cases.
In \cref{sec:tag-space:tag-map}, we detail how to build efficient tag maps using the generalized tags to reduce the amount of unnecessary work done by the system.
Finally, we discuss how to extend our work to accommodate NULLs and three-valued logic in \cref{sec:tag-space:unknown}.

\subsection{Naive Strategy}
\label{sec:tag-space:naive}

In the most naive strategy to tag management, each tag contains true/false assignments to only base \patom[s] (as opposed to the arbitrarily complex predicate expressions described in \cref{sec:model:setup}).
Tags like \ttag{t.year > 2000 = T, mi\_idx.score > '7.0' = F} are valid, while \ttag{t.year > 2000 AND mi\_idx.score > '7.0' = F} are not because it contains an AND.
The tag map for each operator is built as follows.
First, base tagged relations, created directly from the base tables, contain only one \relslice[] with the ``empty'' tag which contains no assignments (i.e, \ttag{}).
For filter operators, create a tag map entry for each input tag and output both the ``true'' and ``false'' results.
More formally, if the filter operator's associated predicate is $P$, for each input tag $I$, create the tag map entry:
\[
  I \rightarrow \{T: I \cup \{P = T\},\; F: I \cup \{P = F\}\}
\]
For join operators, take the full Cartesian product of the input left and right tags.
In other words, for each pairing of left and right input tags $(L, R)$, generate the following tag map entry:
\[
  (L, R) \rightarrow L \cup R
\]
Finally, the projection operator's set of allowed tags is the set of tags whose assignments satisfy the overall predicate expression.

This naive strategy is capable of performing disjunctive predicate pushdown.
Since the predicate evaluation results are stored in tags, filter operators can be pushed down to the base table, and the projection operator simply selects all tags which satisfy the query's overall predicate expression.
However, the naive strategy suffers from two large weaknesses.
%(1) Filter operators output both ``true'' and ``false '' each time, so they do not actually filter any tuples, and the joins end up being performed on the full base tables.
(1) Filter operators output both ``true'' and ``false '' each time, so they do not actually filter any tuples.
Since the join operators' tag maps contain the full Cartesian product of input tags, this results in joins being performed on the full base tables.
(2) In addition, each filter operator outputs two new tags for each input tag, multiplying the number of tags by two.
Thus, after $n$ filter operators, there will be $2^n$ output tags, and this exponential blowup can potentially cause the overhead to outweigh the benefits.

%We now turn our attention to tag management; that is, how should the planner decide which tags to generate, which tags to preserve, and what tags should look like when they are combined.
%More technically, given a certain query plan, what tag maps should the planner build for each operator to minimize the amount of redundant work done by the execution engine.
%Specifically, which tags should be generated, which tags should be preserved, and how should tags be combined?
%This is nontrivial because a naive scheme that simply stores all true/false assignment values of predicates in tags does not actually filter any tuples and can lead to an exponential number of tags (each predicate outputs both ``true'' and ``false'' tags for each input tag, doubling the number of tags in the system after each predicate), causing the overhead of tagged execution to outweigh the benefits.
%To address this problem, we first introduce a technique called tag generalization to reduce the tag space (\cref{sec:tag-space:general}), then we show how planners can use tag generalization to build highly efficient tag maps (\cref{sec:tag-space:tag-map}).

\subsection{Tag Generalization}
\label{sec:tag-space:general}
\begin{figure}
  \centering
  \ifthenelse{\papermode = \submission}{%
    \includegraphics[width=\linewidth]{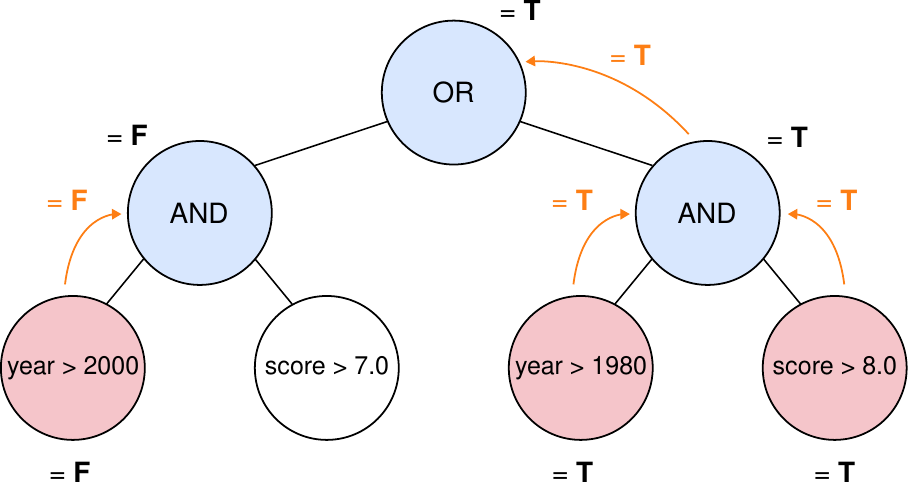}
  }{%
    \includegraphics[width=.7\linewidth]{figs/pred-tree2.pdf}
  }
  \caption{Propagation with predicate tree from \cref{query:ex}.}
  \label{fig:pred-tree}
\end{figure}

To address the weaknesses of the naive strategy, we introduce the idea of generalizing tags.
Tag generalization is a technique which generalizes the assignments in a tag based on the structure present in the query's predicate expression and Boolean implication.
A generalized tag can be used in place of an ungeneralized tag, and multiple ungeneralized tags may generalize to the same generalized tag.
Thus, replacing all the tags in the naive strategy with generalized tags reduces the total number of tags in the system and avoids the exponential blowup in many cases.
%Tag generalization is a technique which groups tags together according to the structure present in the query's predicate expression.
%Generalized tags are used in place of the individual tags they group, and this reduces the total number of tags in the system and the total tag space.
An additional benefit of tag generalization is that it clearly exposes (as early as possible) which tags will not be part of the final output.
The planner can then use this information to build efficient tag maps which discard tuples that do not satisfy the overall predicate expression as early as possible (\cref{sec:tag-space:tag-map}).

In tag generalization, the query's predicate expression is represented as a \emph{predicate tree}, in which each node represents a subexpression.
Leaf nodes represent the base predicates, and intermediate AND/OR/NOT nodes apply their operations to their children to form their subexpressions\footnote{The predicate tree is also normalized such that an intermediate node cannot be of the same type as their parent (i.e., an AND node's parent cannot also be an AND node)}.
%Leaf nodes represent the base predicates, intermediate AND/OR nodes combine children subexpression nodes\footnote{The predicate tree is also normalized such that parents of AND nodes are OR nodes and vice versa, achieving an interleaving of ANDs and ORs across different levels.}, and each intermediate NOT node is applied to its single child.
The root node refers to the entire predicate expression.
%\patom[cs] form the leaf nodes, and intermediate AND/OR nodes combine children subexpression nodes\footnote{The predicate tree is also normalized such that parents of AND nodes are OR nodes and vice versa, achieving an interleaving of ANDs and ORs across different levels.}.
\cref{fig:pred-tree} shows the predicate tree for \cref{query:ex}.

%\srm{The goal of generalization is to assign a tag to each node in this tree.  We start with the tags of each leaf predicate, and propagate tags upwards, eventually creating a tag assignment for each node in the tree.  These generalized tags are then used to construct the {\it tag maps} for each operator; during tag map creation, tags that definitely do not result in query outputs are discarded.  We begin with a description of the generalization process and the show how this is used during tag map generation in \cref{sec:tag-space:tag-map}).}

%A new tag is generated to replace that group of tags, and this decreases the total number of tags in the system, thereby reducing the tag space.
%In our system, tag generalization is performed any time a new tag is created, and the generalized tag is used instead of the original tag.
%Although other Boolean expression minimization techniques may be used, optimal minimization of Boolean expressions is known to be $\sum_{2}^P$-complete~\cite{buchfuhrerComplexityBooleanFormula2011}, so generic techniques may not suffice.

Tag generalization works by propagating a tag's assignments upwards in the predicate tree.
As an assignment travels upwards, it becomes more general and encompasses more possible tags.
%Any time a new tag is created, we attempt to \emph{generalize} that tag's assignments by propagating them upwards in this predicate tree.
%For example, the tag \ttag{t.year > 2000 = F} can be generalized into \ttag{(t.year > 2000 AND mi\_idx.score > '7.0') = F} because \texttt{t.year > 2000 = F} implies \texttt{(t.year > 2000 AND mi\_idx.score > '7.0')} is false according to \cref{query:ex}'s predicate tree.
%The generalized tag \ttag{(t.year > 2000 AND mi\_idx.score > '7.0') = F} can then be used in place of tags such as \ttag{t.year > 2000 = F}, \ttag{t.year > 2000 = F, mi\_idx.score > '7.0' = T}, and \ttag{t.year > 2000 = T, mi\_idx.score > '7.0' = F} because these tags all imply the generalized tag.
For example, the assignment \texttt{t.year > 2000 = F} can be generalized into the assignment \texttt{(t.year > 2000 AND mi\_idx.score > '7.0') = F} because the former implies the latter according to \cref{query:ex}'s predicate tree.
The generalized tag \ttag{(t.year > 2000 AND mi\_idx.score > '7.0') = F} can then be used in place of not only \ttag{t.year > 2000 = F}, but also tags such as \ttag{mi\_idx.score > '7.0' = F} and \ttag{t.year > 2000 = T, mi\_idx.score > '7.0' = F} because these tags also imply the generalized tag.
%A single tag containing only this assignment can then be used to represent multiple tags, such as \ttag{t.year > 2000 = F}, \ttag{t.year > 2000 = F, mi\_idx.score > '7.0' = T}, and \ttag{t.year > 2000 = T, mi\_idx.score > '7.0' = F}.
%may all be combined into a single tag using this generalized assignment (\ttag{t.year > 2000 AND mi\_idx.score > '7.0' = F}), thereby reducing the tag space.

\begin{algorithm}
  \begin{algorithmic}[1]
    \Require Tag ``tag''
    %\Ensure Propagated tag
    %\State fringe $\gets \{(\text{pred}, \text{value}) \in \text{tag}\}$
    %\While{fringe $\ne \varnothing$}
    %\State (pred, value) $\gets$ fringe.pop()
    %\For{parent $\in$ parents(pred)}
    %\If{(type(parent) = AND and value = \False) or (type(parent) = OR and value = \True) or (type(parent) = AND and $\forall c \in$ children(parent), ($c$, \True) $\in$ tag) or (type(parent) = OR and $\forall c \in$ children(parent), ($c$, \False) $\in$ tag)}
    %\State tag $\gets$ tag $\cup$ (parent, value)
    %\State fringe $\gets$ fringe $\cup$ (parent, value)
    %\EndIf
    %\EndFor
    %\EndWhile
    %\State fringe $\gets \{\text{pred} \in \text{keys(tag)}\}$ \label{line:prop-up:main-beg}
    \State fringe $\gets$ keys(tag) \label{line:prop-up:main-beg}
    \While{fringe.isNotEmpty()}
    \State pred $\gets$ fringe.pop()
    \For{parent $\in$ parents(pred)}
    %\If{(isAnd(parent) and tag[pred] = \False) or \\ (isOr(parent) and tag[pred] = \True) or \\ (isAnd(parent) and $\forall c \in$ children(parent), tag[$c$] = \True) or \\ (isOr(parent) and $\forall c \in$ children(parent), tag[$c$] = \False)}
    \If{\canprop(pred, parent, tag)}
    %\If{(isAnd(parent) and tag[pred] = \False) or \\ (isOr(parent) and tag[pred] = \True) or \\ (isAnd(parent) and $\forall c \in$ children(parent), tag[$c$] = \True) or \\ (isOr(parent) and $\forall c \in$ children(parent), tag[$c$] = \False)}
    %\State tag[parent] $\gets$ tag[pred]
    \State \doprop(pred, parent, tag)
    \State fringe.push(parent)
    \EndIf
    \EndFor
    \EndWhile \label{line:prop-up:main-end}
    \State \Return \topmost(root, tag)
    \Statex
    \Function{\canprop}{pred, parent, tag}
    \State \Return isNot(parent) or \\
    \hspace*{.5em} (isOr(parent) and tag[pred] = \True) or \\
    \hspace*{.5em} (isAnd(parent) and tag[pred] = \False) or \\
    \hspace*{.5em} (isOr(parent) and $\forall c \in$ children(parent), tag[$c$] = \False) or \\
    \hspace*{.5em} (isAnd(parent) and $\forall c \in$ children(parent), tag[$c$] = \True)
    \EndFunction
    \Statex
    \Function{\doprop}{pred, parent, tag}
    \If{isNot(parent)}
    \State tag[parent] $\gets$ $\neg$tag[pred]
    \Else
    \State tag[parent] $\gets$ tag[pred]
    \EndIf
    \EndFunction
    \Statex
    \Function{\topmost}{node, tag}
    \If{tag = $\{\}$}
    \State \Return $\{\}$
    \ElsIf{node $\in$ keys(tag)}
    \State \Return $\{\text{node}: \text{tag[node]}\}$
    \Else
    \State \Return $\bigcup_{c \in \text{children(node)}}$ \topmost($c$, tag)
    \EndIf
    \EndFunction
  \end{algorithmic}
  \caption{\propup{}}
  \label{alg:prop-up}
\end{algorithm}

\propup{} from \cref{alg:prop-up}\ifthenelse{\papermode = \cameraready}{ (next page)}{} outlines the general algorithm for generalizing tags.
The input to \propup{} is a tag which maps predicate expressions to true/false values; each key-value pair constitutes an assignment.
At its core, \propup{} is a fringe-based algorithm which propagates a predicate expression's assignment to its parent if the parent's assignment can be implied by its children.
The fringe is initialized with the set of all predicate expressions with an assignment in the input tag (i.e., ``keys(tag)'').
The function \canprop{} checks for the condition of Boolean implication and returns true if one of five conditions is satisfied:
\begin{enumerate*}[(a)]
  \item If the parent is a NOT node.
  \item If the predicate expression's assignment value is true and the parent is an OR node.
  \item If the predicate expression's assignment value is false and the parent is an AND node.
  \item If the parent is an OR node and all its children have false assignments. \label{item:prop-up:cond4}
  \item If the parent is AND node and all its children have true assignments. \label{item:prop-up:cond5}
\end{enumerate*}
The actual propagation is done with \doprop{}, and unless the parent is a NOT node, the predicate expression's assignment value is assigned directly to the parent.
If the propagation is successful, the parent is added to the fringe, and the loop continues.
Note that the same predicate expression may appear multiple times in the predicate tree, so the ``parents'' function returns the parent for each instance.
Once all assignments have been propagated as much as possible, \topmost{} is called on the resulting tag to collect only the topmost assignments.
Assignments to predicate expressions, which are descendants of other predicate expressions with assignments, are all are discarded.
Note that \propup{} runs in $O(n)$ time, in which $n$ is the number of \patom[s].

\cref{fig:pred-tree} shows the predicate tree for \cref{query:ex} undergoing this process for the tag \ttag{t.year > 2000 = F, t.year > 1980 = T, mi\_idx.score > '8.0' = T}.
The red-colored leaf nodes mark the nodes with the initial assignments, the orange arrows show the propagation, and the blue-colored intermediate nodes mark the nodes which get assignments due to propagation.
After \topmost{} is called, only the assignment at the root node remains, and the output of \propup{} is \ttag{(t.year > 2000 AND mi\_idx.score > '7.0') OR (t.year > 1980 = T AND mi\_idx.score > '8.0') = T}.

Using the predicate tree to generalize assignments also has the benefit of making it clear which predicate expressions still need to be applied to which tags.
In the previous example, the tag after propagation \ttag{(t.year > 2000 AND mi\_idx.score > '7.0') OR (t.year > 1980 = T AND mi\_idx.score > '8.0') = T} has a true assignment to the overall expression, signifying that every tuple with this tag should be part of the final output.
Thus, no additional filters need to be run on this tag.
Similarly, if the overall predicate expression has a false assignment, it signifies that every tuple with the tag does not match the criteria for the query, and these tuples can safely be removed from the pipeline.
%\cref{sec:tag-space:tag-map} shows how planners leverage this idea to build efficient tag maps.
%on this idea when generating tag maps.
%builds on this idea to genruses this idea provides more details on how planners can use these benefits.

\textbf{Duplicates.}
As mentioned, the same predicate expression may appear multiple times in the overall predicate expression.
In fact,  conversations with real users of disjunctive workloads have led us to believe that this is actually a common occurrence for disjunctive queries.
Thus, careful consideration has gone into the development of \propup{} and \topmost{} to handle this case.
Specifically, \propup{} allows different levels of propagation for different instances of the same predicate expression, and \topmost{} only removes a predicate expression's assignment if every instance has an ancestor with an assignment.
It is this treatment of predicate expressions which allows tagged execution to evaluate each \patom[] exactly once.

\textbf{Limitations.}
Although generalizing tags reduces the tag space in most cases, the number of tags produced can still be exponential in the worst case.
Consider the predicate expression with the form $(X_1 \lor Y_1) \land (X_2 \lor Y_2) \land ... \land (X_n \lor Y_n)$, in which all $X_i$ and $Y_i$ are \patom[s].
If the filter operators are ordered and applied in $X_1, X_2, ..., X_n,Y_1,Y_2,...,Y_n$  order, then $2^{n}$ tags are still required to keep track of the unique \relslice[s].
However, such degenerate predicate expressions are not so common in practice, and the planner can interleave the $Y_i$ \patom[s] (i.e., use a different plan) to reduce the number of tags.
%In practice, however, we have found this to not be a problem, and the planner can always choose to interleave the $Y_i$ \patom[s] to reduce the tag space.

%Furthermore, this method of generalization ensures that the resulting assignments do not ever have to be split, as long as all filter operators have predicate expressions which correspond to nodes in the predicate tree.
%Assuming the generalized assignment is \texttt{t.year > 2000 AND mi\_idx.score > '7.0' = F}, any filter operators with predicate expressions that are descendants of the generalized assignment, such as \texttt{mi\_idx.score > '7.0' = T}, do not need to be applied to this tag since the results will not changed the generalized assignment value.
%For predicate expressions which are not descendants, such as
%The reason is twofold.
%Given a generalized assignment:
%\begin{enumerate*}
%  \item If a filter operator has a predicate expression that is descendent of that generalized assignment, it need not be applied since the results of the predicate expression will not change the generalized assignment value.
%  \item If a filter operator has a predicate that is not descendent, then the predicate expression must be applied to every tuple with that tag.
%\end{enumerate*}
%For example,
%a filter operator with a predicate expression that is a descendant of the generalized assignment need not be applied, since the result of the predicate expression will not changed the generalized assignment.

\subsection{Building Tag Maps}
\label{sec:tag-space:tag-map}
We now describe how the planner should build the tag maps for each operator in a given query plan.
%\srm{We now describe how tag maps are created that, for a given query plan and predicate tree, determine which tags to retain and which to discard. }
Tag maps should minimize the amount of work performed by operators during runtime while still ensuring correctness of the query plan.
To accomplish this, our planners follow two main precepts when building tag maps:
\begin{enumerate*}[(1)]
  \item Avoid generating tags which do not get used by the rest of the pipeline.
  \item Do not apply filters on tags if it does not help refine the selection process.
\end{enumerate*}
Fortunately, tag generalization make it easy to recognize both cases.

For (1), we can identify such tags if they include a false assignment to the root node of the predicate tree.
Tuples with this tag are guaranteed to not satisfy the overall predicate expression and can be discarded without further consideration.
On the other hand, if an assignment to the root node does not emerge after generalizing a tag, it signifies that a portion of the tuples associated with that tag may still satisfy the overall predicate expression, and more filter operators must be applied to refine the associated \relslice[].
For example, in \cref{fig:query-plan}, the filter operator with the predicate expression \texttt{t.year > 2000} is applied before the filter operator with the predicate expression \texttt{t.year > 1980}.
The first filter operator results in the tags \ttag{t.year > 2000 = T} and \ttag{t.year > 2000 = F}, and after tag generalization, \ttag{t.year > 2000 = F} becomes \ttag{t.year > 2000 AND mi\_idx.score > '7.0' = F}.
However, the \relslice[] with the false tag may still contain tuples that satisfy the overall predicate expression, namely the movies produced after 1980 with a score greater than 8.0.
Thus, the \relslice[] must be kept.
However, after applying the second filter operator to this \relslice[], movies produced before 1980 will generate the tag \ttag{t.year > 2000 AND mi\_idx.score > '7.0' = F, t.year > 1980 = F}, and this generalizes to \ttag{(t.year > 2000 AND mi\_idx.score > '7.0') OR (t.year > 1980 AND mi\_idx.score > '8.0') = F}.
At this point, we can be sure the \relslice[] with this tag does not contain any tuples which satisfy the overall predicate expression, since movies produced before 1980 do not satisfy the overall criteria.
Thus, for the second filter operator, the planner should omit the negative output tag (\negtag{}) from the tag map entry.

For (2), let us assume we apply the filter operators with \texttt{t.year > 2000} and \texttt{mi\_idx.score > '7.0'} in that order after joining the tables.
In this case, applying the predicate expression \texttt{mi\_idx.score > '7.0'} to the \relslice[] with the tag \ttag{t.year > 2000 = F} is pointless.
The tuples with this tag are already guaranteed to not satisfy the first disjunctive clause in \cref{query:ex}, so dividing this set of tuples into those that satisfy/do not satisfy \texttt{mi\_idx.score > '7.0'} does nothing to help the tuple selection process.
%so they must be tested for the second disjunctive clause, and testing them further for \ttag{t.year > 2000 = F} would.
%the predicate expressions \texttt{t.year > 1980} and \texttt{mi\_idx.score > '8.0'} must be applied to this \relslice[] anyway.
%More explicitly, dividing the \relslice[] with the tag \ttag{t.year > 2000 = F} into the tuples that satisfy/do not satisfy \texttt{mi\_idx.score > '7.0'} does nothing to help the tuple selection process, when \texttt{mi\_idx.score > '8.0'} must be applied later anyway.
Precept (2) states that such cases should be avoided, and generalized assignments make it trivial to identify such cases.
The tag \ttag{t.year > 2000 = F} after generalization becomes \ttag{t.year > 2000 AND mi\_idx.score > '7.0' = F}, and this predicate expression is an ancestor (in the predicate tree) of the predicate expression we were trying to apply (\texttt{mi\_idx.score > '7.0'}).
Thus, both positive and negative outcomes of \texttt{mi\_idx.score > '7.0'} (i.e., \ttag{t.year > 2000 AND mi\_idx.score > '7.0' = F, mi\_idx.score > '7.0' = T} and \ttag{t.year > 2000 AND mi\_idx.score > '7.0' = F, mi\_idx.score > '7.0' = F}) reduce to the same tag \ttag{t.year > 2000 AND mi\_idx.score > '7.0' = F} after generalization, signifying that the application of \texttt{mi\_idx.score > '7.0'} to this tag was not very helpful in refining the tuples according to overall predicate expression.
%The outcome of \texttt{mi\_idx.score > '7.0'} cannot the assignment to the ancestor predicate expression, so we should avoid applying \texttt{mi\_idx.score > '7.0'} to this tag.
%More explicitly, both \ttag{t.year > 2000 AND mi\_idx.score > '7.0' = F, mi\_idx.score > '7.0' = T} and \ttag{t.year > 2000 AND mi\_idx.score > '7.0' = F, mi\_idx.score > '7.0' = F} simplify to \ttag{t.year > 2000 AND mi\_idx.score > '7.0' = F} after running \propup{} on them.

With these precepts in mind, we now describe the exact process our planners follow for building tag maps.
The overall construction is similar to the naive strategy, except it includes tag generalization and the two precepts.
First, base tagged relations, created directly from the base tables, contain only one \relslice[] with the ``empty'' tag which contains no assignments (i.e, \ttag{}).
%Note that base tagged relations all start with only the empty tag (i.e., the tag with no assignments).

\textbf{Filter.}
For filter operators, if the associated predicate expression is $P$, for each input tag $I$:
\begin{enumerate}[nosep]
  \item
    If each instance of $P$ in the predicate tree has an ancestor with an assignment in $I$, do nothing (Precept (2)).
  \item
    %Otherwise, create an entry in the tag map for $I$.
    Otherwise, run \propup{} on both positive and negative output tags:
    \begin{align*}
      P & \gets \propup(I \cup \{P = T\}) \\
      N & \gets \propup(I \cup \{P = F\})
    \end{align*}
    %\begin{itemize}
    %  \item $\postag  \gets \propup(T \cup \{P = \ttTrue\})$
    %  \item $\negtag  \gets \propup(T \cup \{P = \ttFalse\})$
    %\end{itemize}
    And create a tag map entry for $I$:
    \[
      I \rightarrow \{T: P,\; F: N\}
    \]
    If either $P$ or $N$ include a false assignment to the root node, remove it as an output tag from the entry (Precept (1)).
\end{enumerate}
%In the first case, the output of $P$ cannot affect the assignments in $T$, so applying $P$ to $T$ has no meaning.
%Thus, $T$ is simply passed along with no entry in the tag map.
%In the second case, both positive and negative tags are included in the output unless the generalization process can determine that the result is a tag which does not satisfy the overall predicate expression.
%Tuples with such a tag do not need to be further processed and can be removed from consideration.

\textbf{Join.}
For join operators, take the full Cartesian product of the input left and right tags.
For each pairing of left and right input tags $(L, R)$, generate the output tag by generalizing the union:
\[
  O \gets \propup(L \cup R)
  %\outtag \gets \propup(\lefttag{} \cup \righttag)
\]
If $O$ does not have a false assignment to the root node (Precept (1)), create an entry in the tag map:% with the appropriate \lefttag{}, \righttag{}, and \outtag{}.
\[
  (L, R) \rightarrow O
\]

\textbf{Projection.}
For projections operators, restrict the set of allowed tags to only the tag with a true assignment to the root node.

\subsection{Extension to Three-Valued Logic}
\label{sec:tag-space:unknown}
Standard SQL allows for NULL values, and evaluating a \patom[] on a NULL value often results in a non-true/false, ternary value called ``unknown''.
Fortunately, our framework extends very naturally to this three-valued logic, and only four changes need to be made:
\begin{enumerate}
  \item Instead of restricting a tag's assignments to either be true/false, a tag's assignments may now be one of true/false/unknown.
  \item For filter operator tag map entries, include an optional output \unktag{} for unknown results.
  \item The functions \canprop{} and \doprop{} in \cref{alg:prop-up} must be updated to handle unknown values.
    %replaced with \canpropnull{} and \dopropnull{} from \cref{alg:prop-up-null} to handle unknown values.
    %\canpropnull{} is almost the same except that conditions 4 and 5 accept unknown assignments to children, meanwhile
    %be updated to handle unknown values.
    For \canprop{}, the only difference is for conditions \ref{item:prop-up:cond4} and \ref{item:prop-up:cond5}, which must now be changed to:
    \ref{item:prop-up:cond4} If the parent is an OR node and all its children have false or unknown assignments.
    \ref{item:prop-up:cond5} If the parent is AND node and all its children have true or unknown assignments.
    %The function \canprop{} only differs slightly and returns true if one of the following holds:
    %\begin{enumerate*}
    %  \item If the parent is a NOT node.
    %  \item If the predicate expression's assignment value is true and the parent is an OR node.
    %  \item If the predicate expression's assignment value is false and the parent is an AND node.
    %  \item If the parent is an OR node and all its children have false or unknown assignments.
    %  \item If the parent is AND node and all its children have true or unknown assignments.
    %\end{enumerate*}
    On the other hand, \doprop{} must now update the parent's assignment value based on the three-valued logic from the SQL standard~\cite{meltonUnderstandingNewSQL1993} (e.g., false OR unknown $\rightarrow$ unknown).
  \item Finally, applications of Precept (1) must be changed so that tag map entries do not contain output tags with either a false or unknown assignment to the root node.
\end{enumerate}

%\begin{algorithm}
%  \begin{algorithmic}[1]
%    \Function{\canpropnull}{pred, parent, tag}
%    \State \Return isNot(parent) or \\
%    \hspace*{.5em} (isOr(parent) and tag[pred] = \True) or \\
%    \hspace*{.5em} (isAnd(parent) and tag[pred] = \False) or \\
%    \hspace*{.5em} (isOr(parent) and $\forall c \in$ children(parent), tag[$c$] = \False/\Unknown) or \\
%    \hspace*{.5em} (isAnd(parent) and $\forall c \in$ children(parent), tag[$c$] = \True/\Unknown)
%    \EndFunction
%    \Statex
%    \Function{\dopropnull}{pred, parent, tag}
%    \If{isAnd(parent)}
%    \State tag[parent] $\gets$ $\bigwedge_{c\in \text{children(parent)}}$ tag[$c$]
%    \ElsIf{isOr(parent)}
%    \State tag[parent] $\gets$ $\bigvee_{c\in \text{children(parent)}}$ tag[$c$]
%    \Else
%    \State tag[parent] $\gets$ $\neg$ tag[pred]
%    \EndIf
%    \EndFunction
%  \end{algorithmic}
%  \caption{\canpropnull{} and \dopropnull{}}
%  \label{alg:prop-up-null}
%\end{algorithm}

\section{Planning}
\label{sec:plan}

Now that we understand how to execute queries under tagged execution, and how to build efficient tag maps which effectively prune tuples from disjunctive queries early, we describe how we can generate plans that take advantage of these early filtering capabilities.
%$to minimize the amount of redundant work done by the execution engine, we describe how we can generate plans to take advantage of these capabilities.
%, for a given query plan, we can build tag maps which effectively prune tuples from disjunctive queries early in plan execution.  We also describe how we can generate plans that take advantage of these early filtering capabilities.
We begin by presenting cost models for tagged execution (\cref{sec:plan:cost}) and then give the details for several planners (\cref{sec:plan:planner}).
Note that it is not the goal of this work to produce the most advanced, optimal planner for tagged execution.
Rather, we  present a few simple planners which highlight the advantages of tagged execution and demonstrate that even these simple planners can obtain substantial performance compared to existing methods.

\subsection{Cost Models}
\label{sec:plan:cost}
Although tagged execution introduces a new model for query execution, internally, tagged operators employ the same relational operators from traditional query execution.
The only difference is that the input changes from whole relation(s) to individual \relslice[](s).
As such, our suggested cost models mirror existing models quite closely, differing mainly in that our cost models are summations of the costs of individual  \relslice[s].

For filter operators, the cost model is:
\[
  C_{filter} = \alpha \sum_{I \in M} F_P \left|R[ I ]  \right|
\]
Here, $C_{filter}$ measures the cost of applying the filter operator with predicate expression $P$ on input tagged relation $R$.
The expression $I \in M$ iterates over each input tag in the given tag map, and $F_P$ is a certain constant cost factor specific to $P$.
The expression $R[I]$ selects the \relslice[] associated with tag $I$ in $R$, and $|R[I]|$ measures the cardinality of the selected \relslice[].
Finally, $\alpha$ is a constant cost factor used to calibrate filter costs with respect to join costs.
In short, the total cost is the summation of the costs from applying predicate expression $P$ to every \relslice[] whose tag appears as an input in the given tag map.

For join operators, the cost model is:
\begin{align*}
  C_{join} &= C_{hash\_build} + C_{hash\_lookup} + C_{index\_build} \\[1em]
  C_{hash\_build} &= F_{hash\_lookup} \left| R_{left}' \right| + F_{hash\_build} \cdot \text{unique}(R_{left}') \\
  C_{hash\_lookup} &= F_{hash\_lookup} \left| R_{right}' \right| \\
  C_{index\_build} &= F_{index\_build} \left| R_{left}' \Join R_{right}' \right|
  %\left( | \text{tables}(R_{left})| + |\text{tables}(R_{right}) | \right)
\end{align*}
As mentioned, all join operators in our system are implemented as hash joins, so the cost of a join can be split into three components:
\begin{enumerate*}
  \item $C_{hash\_build}$ -- the cost of building a hash map from the left input tagged relation\footnote{Although this cost estimate assumes the hash map will be built from the left side, our system actually makes an estimate from both sides and chooses the cheaper side.},
  \item $C_{hash\_lookup}$ -- the cost of performing the hash lookups from the right input tagged relation, and
  \item $C_{index\_build}$ -- the cost of building the index for the output joined tagged relation.
\end{enumerate*}
Building the hash map requires performing a hash lookup for each element and creating entries for each unique element.
Accordingly, $F_{hash\_lookup}$ is some cost factor associated with hash lookups, $|R_{left}'|$ measures the cardinality of $R_{left}'$, $F_{hash\_build}$ is some cost factor associated with creating hash map entries, and $\text{unique}(R_{left}')$ counts the number of unique elements in $R_{left}'$.
%Note $F_{hash\_lookup}$ and $F_{hash\_build}$ are cost factors that depend on the size of the element being hashed, so they depend on the join constraint.
Note that $R_{left}'$ is not the left input tagged relation; rather, $R_{left}'$ is the union of \relslice[s] in the left input tagged relation whose tags have at least one matching entry in the given tag map; $R_{right}'$ from $C_{hash\_lookup}$ is defined similarly.
%Similarly, $R_{right}'$ from $C_{hash\_lookup}$ is the union of \relslice[s] in the right input tagged relation whose tags have at least one matching entry in the given tag map.
Finally, the cost of building the index for the output joined tagged relation is the product of some cost factor $F_{index\_build}$ and its cardinality.
%The cost factor $F_{index\_build}$ depends on the number of tables in the output.

Note that the cost models for both filter and join operators require cardinality estimates for tagged relations/\relslice[s].
In general, the cardinality estimate for a tagged relation is given as the sum of its \relslice[s]'s cardinality estimates.
For filters, we measure and use the selectivities of \patom[s] along with the independence assumption.
For joins, we use PostgreSQL's cardinality estimations of joins~\cite{leisHowGoodAre2015}.

%Note that this estimate is different from estimating based on each individual \relslice[] join. Although, that may have also sufficed, this was closer to what the system implements, so we use this.

\subsection{Planners}
\label{sec:plan:planner}
Here, we present several planners for tagged execution.
Each planner optimizes for a different situation, and in our system, we use the \combined{} planner, which estimates the cost of the plan produced by each of the following planners and selects the cheapest one.

\pushdown{}.
This planner creates a filter operator for each \patom[] in the predicate expression and pushes all filter operators down to the base table level.
All joins are performed after the filter operators and are ordered greedily; whichever join would produce the smallest cardinality tagged relation is performed next (this is actually the join ordering used for all our planners).
After pushdown, if there are multiple filter operators for a single table, they are sorted in \emph{benefiting} order.
%\srm{I know we don't want to give the details of the benefit algorithm but perhaps we give describe the benefit order of a particular set of predicates under the cost model?}
The benefit score of a filter operator is calculated with respect to a set of filter operators, and it  estimates the benefit of applying that filter operator first before the set of filter operators.
The score is used here and in other planners to avoid materializing every alternative plan and serves as an effective proxy for plan cost.
The exact method to calculate the benefit score can be found in \ifthenelse{\papermode = \techreport}{\cref{sec:benefit}.}{our technical \ifthenelse{\papermode = \cameraready}{report~\cite{kimOptimizingDisjunctiveQueries2024}.}{report~\cite{OptimizingDisjunctiveQueries2023}.}}
\pushdown{} is the ``naive'' planner which simply pushes down all all filter operators down to the base table level.
It serves as a good baseline planner, since for many queries, the joined relation is much larger than the base table relations, and pushing filter operators down to the base table level prunes as many tuples as early as possible.
\cref{fig:query-plan} serves as an example query plan that \pushdown{} might produce.

\begin{algorithm}
  \begin{algorithmic}[1]
    \State (best\_plan, best\_cost) $\gets$ \pushdown{}()
    \For{filter $\in$ filters(plan)}
    %\State (cmp\_plan, cmp\_cost) $\gets$ (best\_plan, best\_cost)
    \State new\_plan $\gets$ best\_plan
    \While{can\_pullup(new\_plan, filter)}
    \State (new\_plan, new\_cost) $\gets$ pullup\_node(new\_plan, filter)
    \If{new\_cost < best\_cost}
    \State (best\_plan, best\_cost) $\gets$ (new\_plan, new\_cost)
    \EndIf
    \EndWhile
    \EndFor
    \State \Return best\_plan
  \end{algorithmic}
  \caption{\pullup{}}
  \label{alg:pullup}
\end{algorithm}
\pullup{}.
The pseudocode for this planner is presented in \cref{alg:pullup}.
This planner uses the plan produced by \pushdown{}, in which all filter operators are pushed down, as the base plan.
Then, for each filter operator (in reverse benefiting order), the planner considers pulling up~\cite{hellersteinPredicateMigrationOptimizing1993b} the operator by one node in the query plan, and if the resulting plan is cheaper, that plan is used as the base plan from then on.
In the end, the planner returns the cheapest plan it finds.
\pullup{} builds on \pushdown{} for situations in which certain predicate subexpressions are very selective and can cause the joined result to be much smaller than the base tables.
In these cases, it may be cheaper to delay the application of other more expensive subexpressions, so \pullup{} pulls up each filter node to check.
For example, consider the predicate expression: \texttt{(mi\_idx.score = '9.2' OR mi\_idx.score = '9.3') AND t.title ILIKE '\%godfather\%'}.
There are only two movies in the IMDB dataset with a score above 9.0, so applying the score predicates to the \texttt{movie\_info\_idx} table and joining it with the \texttt{title} table will lead to a joined relation of two tuples.
As such applying the expensive regex pattern matching predicate on this joined result may result in a cheaper plan than applying it directly to the entire \texttt{title} table.
\pullup{} can check for situations such as this and output the following plan:
%applying the expensive regex pattern matching predicate \texttt{t.title ILIKE '\%godfather\%'} on the joined result may result in a cheaper plan than applying the predicate to the entire \texttt{title} table.
\ifthenelse{\papermode = \submission}{%
\smallskip
\dirtree{%
  .1 Filter(t.title ILIKE '\%godfather\%').
  .2 Join(t.id = mi\_idx.movie\_id).
  .3 Table(title as t).
  .3 Filter(mi\_idx.score = '9.2').
  .4 Filter(mi\_idx.score = '9.3').
  .5 Table(movie\_info\_idx as mi\_idx).
}
\smallskip
}{%
  \begin{center}
  \begin{varwidth}{.6\linewidth}
\dirtree{%
  .1 Filter(t.title ILIKE '\%godfather\%').
  .2 Join(t.id = mi\_idx.movie\_id).
  .3 Table(title as t).
  .3 Filter(mi\_idx.score = '9.2').
  .4 Filter(mi\_idx.score = '9.3').
  .5 Table(movie\_info\_idx as mi\_idx).
}
\end{varwidth}
\end{center}
}

\iterpush{}.
This planner operates in the ``opposite'' direction of \pullup{}.
The base plan performs all joins first, then all filters are applied in benefiting order.
For each filter operator (in benefiting order), the planner considers pushing down the operator to the base table level.
If the resulting plan is cheaper, that plan is used as the base plan from then on.
In the end, the planner returns the cheapest plan it finds.
One weakness of \pullup{} is that it only considers moving one filter operator at a time.
Sometimes, there is no benefit to moving a single operator; only when multiple filter operators are moved can a change in the plan cost be observed.
Thus, \pullup{} does not always find the optimal plan, and \iterpush{} serves as the opposite extreme, in which the default for all filter operators is to be performed after all the joins, and filter operators are only pushed down if it leads to a cheaper plan.
To take an example, let us consider the predicate expression: \texttt{mi\_idx.score > '9.0' AND (t.title ILIKE '\%godfather\%' OR t.title ILIKE '\%lord\%')}.
Once again, we would like to perform the expensive regex pattern matching predicates on the joined result, since \texttt{mi\_idx.score > '9.0'} is so selective.
Consider the plan in which only \texttt{t.title ILIKE '\%godfather\%'} is pulled up:
\ifthenelse{\papermode = \submission}{%
\smallskip
\dirtree{%
  .1 Filter(t.title ILIKE '\%godfather\%').
  .2 Join(t.id = mi\_idx.movie\_id).
  .3 Filter(t.title ILIKE '\%lord\%') \hspace{0.2em} $\gets$ (1).
  .4 Table(title as t).
  .3 Filter(mi\_idx.score > '9.0').
  .4 Table(movie\_info\_idx as mi\_idx).
}
\smallskip
}{%
  \begin{center}
    \begin{varwidth}{.6\linewidth}
\dirtree{%
  .1 Filter(t.title ILIKE '\%godfather\%').
  .2 Join(t.id = mi\_idx.movie\_id).
  .3 Filter(t.title ILIKE '\%lord\%') \hspace{0.2em} $\gets$ (1).
  .4 Table(title as t).
  .3 Filter(mi\_idx.score > '9.0').
  .4 Table(movie\_info\_idx as mi\_idx).
}
\end{varwidth}
\end{center}
}
\noindent
This plan is not any cheaper compared to the plan in which all filters are pushed down.
This is because the tagged relation at (1) after the filter must include both \ttag{t.title ILIKE '\%lord\%' = T} and \ttag{t.title ILIKE '\%lord\%' = F}, so the join operator ends up joining to the entire \texttt{title} table.
In comparison, the base plan applies both parts of the disjunct to \texttt{title}, and only the tuples with the tag \ttag{t.title ILIKE '\%godfather\%' OR t.title ILIKE '\%lord\%' = T} are joined.
\pullup{} is incapable of recognizing that pulling up both regex predicates would lead to a more optimal plan.
Thus to provide another perspective for these situations, \iterpush{} starts with all filters pulled up, and in this case pushes down only the \texttt{mi\_idx.score > '9.0'} predicate to arrive at the following plan:
\ifthenelse{\papermode = \submission}{%
\smallskip
\dirtree{%
  .1 Filter(t.title ILIKE '\%godfather\%').
  .2 Filter(t.title ILIKE '\%lord\%').
  .3 Join(t.id = mi\_idx.movie\_id).
  .4 Table(title as t).
  .4 Filter(mi\_idx.score > '9.0').
  .5 Table(movie\_info\_idx as mi\_idx).
}
\smallskip
}{%
  \begin{center}
    \begin{varwidth}{.6\linewidth}
\dirtree{%
  .1 Filter(t.title ILIKE '\%godfather\%').
  .2 Filter(t.title ILIKE '\%lord\%').
  .3 Join(t.id = mi\_idx.movie\_id).
  .4 Table(title as t).
  .4 Filter(mi\_idx.score > '9.0').
  .5 Table(movie\_info\_idx as mi\_idx).
}
    \end{varwidth}
  \end{center}
}

\pushand{}.
This planner mimics what existing planners might do for disjunctive predicates if the root node of the predicate tree is an AND node.
Children of the root node which are individual \patom[s] or predicate expressions whose descendent \patom[s] all apply to the same table are pushed down.
The remaining children are performed in increasing order of selectivity after all the joins.
In short, the planner pushes down the conjunctive filter operators it can, and resolves the remaining after the joins.
This planner mostly serves as a comparison point to traditional query execution.
Although it would arrive at the same plan as \pushdown{} for the previous example (in \iterpush{}), it would not be able to perform any pushdown optimizations for simple CNFs of the form $(P_1 \lor P_2) \land (P_3 \lor P_4)$, if $P_1$ and $P_2$ refer to different tables and $P_3$ and $P_4$ refer to different tables.

\textbf{Discussion.}
Note that in all these planners, we could reorder and push down/pull up \patom[s] however we wanted, without having to worry about how it affects the evaluation of the query's predicate expression.
This is one of the highlights of the tagged execution abstraction model.
By encapsulating all the state required to evaluate the predicate expression into tags, tagged execution is able to disentangle the complexities of evaluating a disjunctive predicate expression from the query plan.
As a result, planners for tagged execution can rearrange \patom[s] freely, almost as if they are planning for a predicate expression with only conjunctions.
Not only does this provide a cleaner interface, it also reduces the query plan space, allowing our planners to complete faster.

%Overall, \pushdown{} is the ``naive'' planner which simply pushes every filter operator down to the base table level.
%\pullup{} builds on \pushdown{} for situations in which certain predicate subexpressions are very selective, and the jjhoined tagged relation is much smaller than the base tables.
%In these cases, it may be cheaper to delay the application of other more expensive subexpressions, so \pullup{} attempts to pull up one filter node at a time to check.
%\iterpush{} takes the opposite perspective; the default for all filter operators is to be performed after all the joins, and filter operators are only pushed down if it leads to a cheaper plan.
%Note that the weakness of both \pullup{} and \iterpush{} is that these planners only consider moving one filter operator at a time.
%Sometimes, there is no benefit to moving a single operator; only when multiple filter operators are moved can a change in the plan cost be observed.
%Thus, \pullup{} and \iterpush{} do not always derive the same plan and instead represent opposite extremes, in which the default for filter operators is all pushed down vs all pulled up.
%\combined{} considers the plans produced by both these planners, and even without considering the simultaneous movement of multiple filter operators, we find that it obtains substantial performance benefits over traditional query execution in our evaluation.
%Finally, \pushand{} is included to serve as a comparison point for traditional query execution and measure the overhead of tagged execution.

\section{Evaluation}
\label{sec:eval}

For our evaluation, we wish to compare the runtime performance of the tagged execution model using our planners against the traditional execution model using existing planners.
In addition, we wish to measure the ``overhead'' of executing queries under the tagged execution model and explore how different query parameters can affect the level of benefit that the tagged execution model has to offer.
To this end, we executed queries from the Join Order Benchmark (JOB)~\cite{leisHowGoodAre2015} (\cref{sec:eval:job}) and a set of synthetic experiments varying different query parameters (\cref{sec:eval:synth}) under both tagged and traditional execution models and measured their total runtimes.
%we used both the tagged and traditional execution models to on the Join Order Benchmark (JOB)~\cite{leisHowGoodAre2015} and a set of synthetic experiments varying different query parameters.
%In these experiments, we compared the total runtimes of executing queries using the tagged execution model versus the traditional execution model.
%To this end, in addition to the tagged execution planners from \ref{sec:plan}, we implemented the following two planners representative of traditional executors: \bdisj{} and \bpushand{}:
To represent traditional execution, we implemented the following two planners:
\begin{enumerate}[nosep]
    \item
\bdisj{} is for predicate expressions with OR root nodes in the corresponding predicate trees (e.g., DNFs).
The planner treats each \emph{\clause[]} (i.e., child of the root node) as a separate query; each \clause[] is executed independently of the others, applying pushdown optimizations wherever possible, and a final union operator to combine all the results.
This (or a variation) is the approach taken by many academic papers~\cite{straubeQueriesQueryProcessing1990}~\cite{jarkematthiasQueryOptimizationDatabase1984} \cite{muralikrishnaOptimizationMultipledisjunctQueries1988}~\cite{changOptimizationDisjunctiveQueries1997}, and experts recommend rewriting SQL queries to achieve this manually for systems which do not support this internally~\cite{albeAvoidBetterPostgreSQL2018}.
%This (or a similar is the approach taken by many academic papers~\cite{straubeQueriesQueryProcessing1990,jarkematthiasQueryOptimizationDatabase1984,muralikrishnaOptimizationMultipledisjunctQueries1988,changOptimizationDisjunctiveQueries1997},
%Many real systems employ this strategy for predicate expressions with disjunctions, and for systems which do not perform this internally, experts recommend rewriting the SQL query to do so explicitly~\cite{albeAvoidBetterPostgreSQL2018}.
\item
\bpushand{} is for predicate expressions with AND root nodes (e.g., CNFs) and is the counterpart to \pushand{} for traditional execution.
Every \clause[] whose \patom[] descendants all refers to the same table are pushed down to that table, and the remaining \clause[s] are performed after all the joins.
This (or a variation) is the approach taken by most real-world systems, such as PostgreSQL~\cite{stonebrakerDesignPostgres1986}, Hyrise~\cite{grundHYRISEMainMemory2010}, and Vertica~\cite{lambVerticaAnalyticDatabase2012}.
\end{enumerate}
Note that similar to the tagged execution planners from \cref{sec:plan}, both \bdisj{} and \bpushand{} order joins greedily.
We compared these traditional execution planners against our combined tagged execution planner \combined{} with its subplanners \pushdown{}, \pullup{}, \iterpush{}, and \pushand{}.

\textbf{System.}
All experiments were conducted on our system \system{}\footnote{\url{https://github.com/alkim0/tagexec}}, a column-oriented database system capable of performing both traditional and tagged query execution.
\system{} is coded in $\sim$12.6k lines of Rust, and data is stored on disk.
When the data for a \relslice[] is needed, \system{} consults the corresponding bitmap, and reads are done using direct I/O calls with a LFU page cache sitting in the middle.
For bitmaps with low selectivity (i.e., only a few values need to be read), only the relevant pages are read from disk.
However, doing the same for bitmaps with high selectivity (i.e., many values need to be read) can lead to substantial penalties from the random I/O, so for all bitmaps with a selectivity above a certain threshold, \system{} instead reads the entire column sequentially, and values are selected in memory.
For the experiments, we ran \system{} on a server running Arch Linux with 40 Intel(R) Xeon(R) Gold 6230 CPU @ 2.10GHz processors, 128GB of memory, and a SSD with 6Gbps of I/O throughput.
%\system{} is capable of executing queries in both tagged and traditional execution models using the aforementioned planners.
%\system{} implements all tagged execution planners introduced in \cref{sec:plan}, as well as a couple traditional execution planners mimicking what real systems do for disjunctive queries.
%The hardware for our experiments was a server running Arch Linux with 40 Intel(R) Xeon(R) Gold 6230 CPU @ 2.10GHz processors, 128GB of memory, and a SSD with 6Gbps of I/O throughput.

%\cref{sec:eval:job} presents our results for JOB, in which we demonstrate the large performance gains of tagged execution for a realistic workload and measure the overhead of executing a query under the tagged execution model.
%\cref{sec:eval:synth} varies different query parameters in a synthetic workload and measures how they affect the performance of tagged execution.

\subsection{Join Order Benchmark}
\label{sec:eval:job}

\ifthenelse{\papermode = \submission}{%
\begin{figure*}
  \centering
  \begin{subfigure}{0.49\linewidth}
    \includegraphics[width=\linewidth]{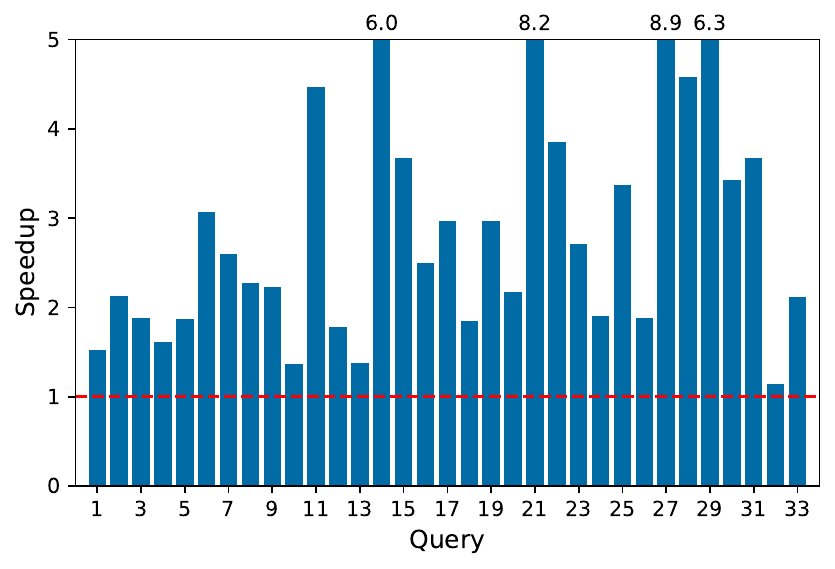}
    \caption{\bdisj{}/\combined{} ($>1$ = tagged execution is better)}
    \label{fig:job-bdisj-combined}
  \end{subfigure}
  \begin{subfigure}{0.49\linewidth}
    \includegraphics[width=\linewidth]{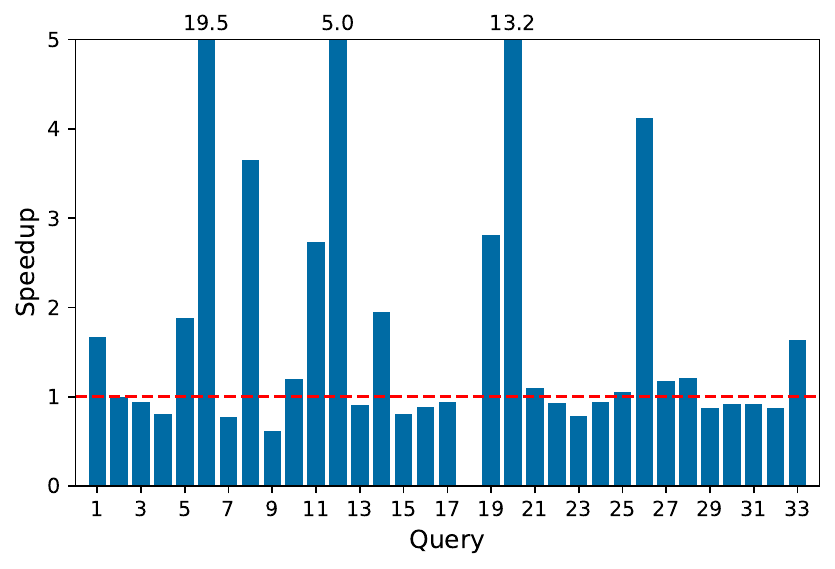}
    \caption{\bpushand{}/\combined{} ($>1$ = tagged execution is better)}
    \label{fig:job-bpushand-combined}
  \end{subfigure}
  \begin{subfigure}{0.49\linewidth}
    \includegraphics[width=\linewidth]{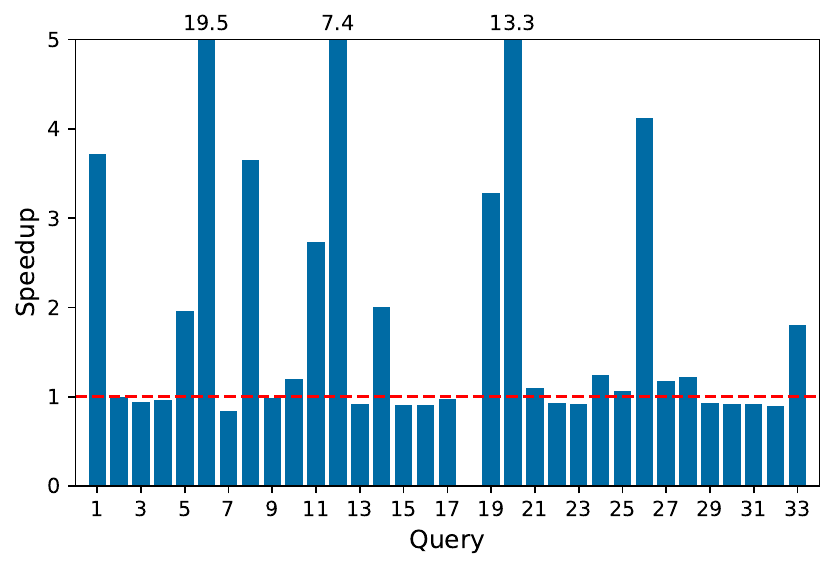}
    \caption{\bpushand{}/\tmin{} ($>1$ = tagged execution is better)}
    \label{fig:job-bpushand-min}
  \end{subfigure}
  \begin{subfigure}{0.49\linewidth}
    \includegraphics[width=\linewidth]{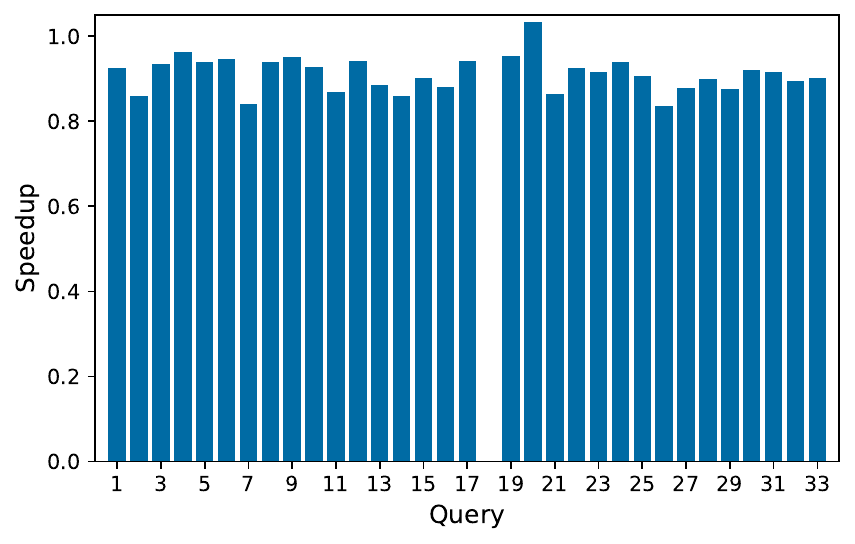}
    \caption{\bpushand{}/\pushand{}  ($>1$ = tagged execution is better)}
    \label{fig:job-overhead}
  \end{subfigure}
  \caption{Speedups in total runtime from using tagged execution for JOB queries.}
  %\caption{Potential speedups of total runtime and overhead of using tagged execution for JOB queries.}
  \label{fig:job}
\end{figure*}
}{%
  \ifthenelse{\papermode = \techreport}{%
    \begin{figure*}
      \centering
      \begin{subfigure}{0.49\linewidth}
        \includegraphics[width=\linewidth]{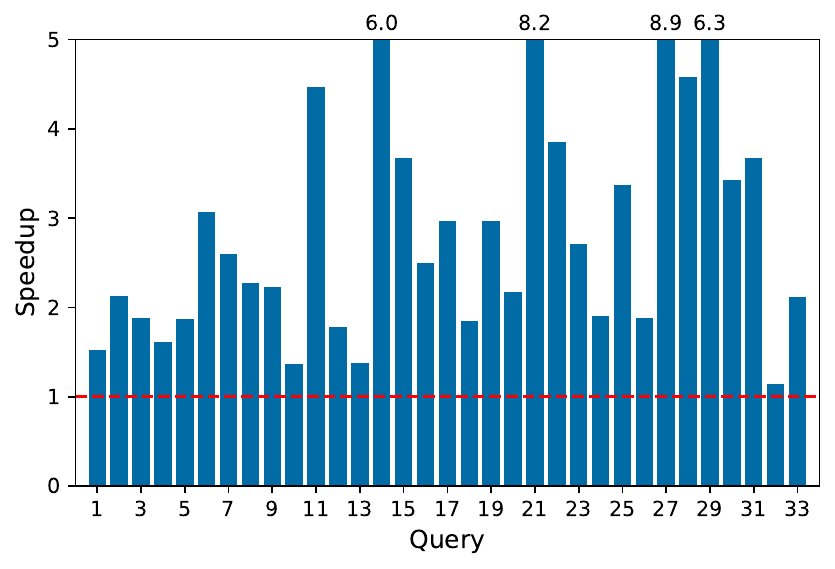}
        \caption{\bdisj{}/\combined{}}
        \label{fig:job-bdisj-combined}
      \end{subfigure}
      \begin{subfigure}{0.49\linewidth}
        \includegraphics[width=\linewidth]{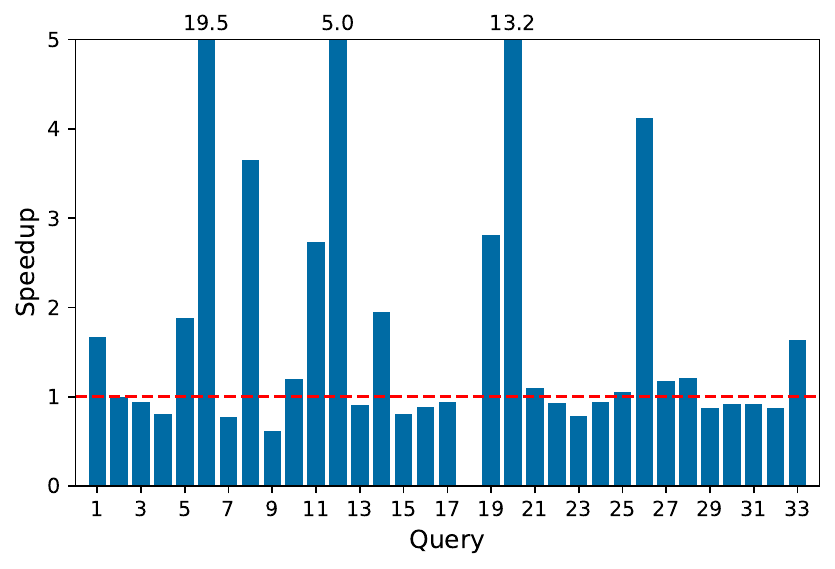}
        \caption{\bpushand{}/\combined{}}
        \label{fig:job-bpushand-combined}
      \end{subfigure}
      \begin{subfigure}{0.49\linewidth}
        \includegraphics[width=\linewidth]{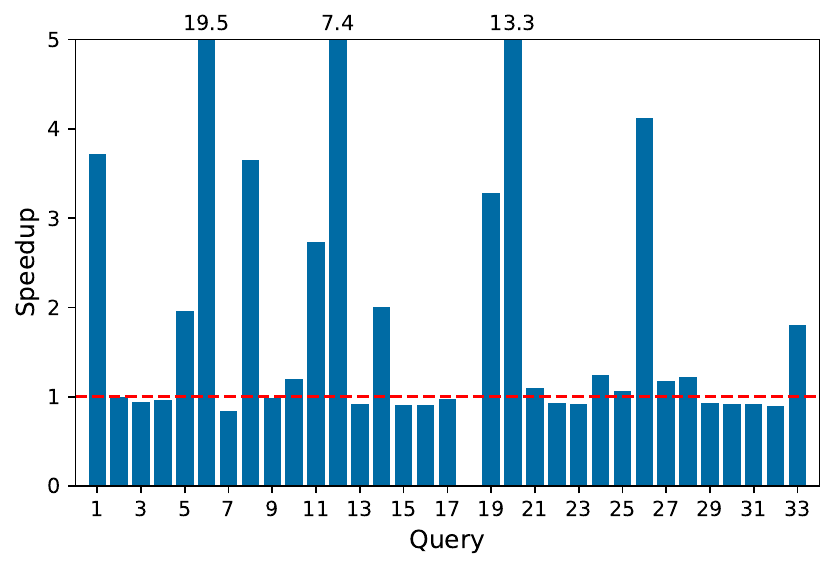}
        \caption{\bpushand{}/\tmin{}}
        \label{fig:job-bpushand-min}
      \end{subfigure}
      \begin{subfigure}{0.49\linewidth}
        \includegraphics[width=\linewidth]{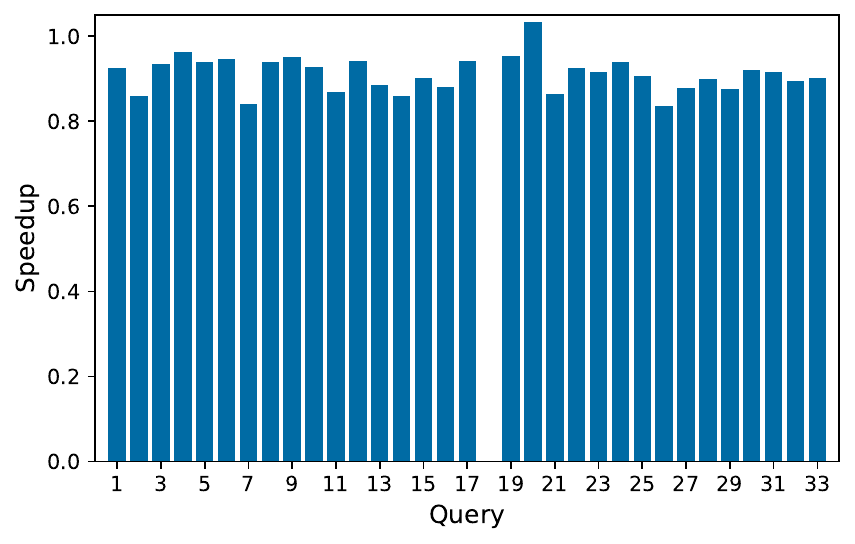}
        \caption{\bpushand{}/\pushand{}}
        \label{fig:job-overhead}
      \end{subfigure}
      \caption{Speedups in total runtime from using tagged execution for JOB queries ($> 1$ means tagged execution is better).}
      %\caption{Potential speedups of total runtime and overhead of using tagged execution for JOB queries.}
      \label{fig:job}
    \end{figure*}
  }{}
}

It should be noted that there is a distinct lack of publicly available disjunctive query workloads.
Although real workloads containing disjunctions have been reported~\cite{fontouraEfficientlyEvaluatingComplex2010}~\cite{jiAtreeDynamicData2021}, their proprietary nature prevents us from accessing them.
%However, we have talked with several owners of proprietary workloads, and they have assured us disjunctions definitely do appear in real workloads and continue to pose problems.
Luckily, JOB provides an avenue to concoct some realistic disjunctive queries.
Vanilla JOB does include a few queries with disjunctions, but these disjunctions never span more than a single table and are not fit for our workload by themselves.
JOB sorts its queries into 33 distinct query groups.
Each query group follows some sort of theme, and more importantly, all queries within a group operate on the same set of tables and use the same join conditions, differing only in the filtering predicate expressions.
Thus, the queries in each query group can be combined together by taking the disjunction of their predicate expressions.
%(e.g., combining a query with predicate expression A and a query with predicate expression B gives a query with predicate expression A OR B).
For example, query group 20 is about superheroes; query 20a searches for superhero movies produced after 1950 with a character named "Iron Man", and query 20c searches for superhero movies produced after 2000 with just any character with the word "Man" in their name\footnote{The actual queries are more complex, but this serves as a summary.}.
Combining queries 20a and 20c would give us one query which searches for superhero movies either produced after 1950 with a character named "Iron Man" or produced after 2000 with any character with the word "Man" in their name.
Doing this for every query group provides us with 33 complex and realistic disjunctive queries to use for evaluation.

%\subsubsection{Planners}
%In addition to the tagged execution planners from \cref{sec:plan}, \system{} implements two planners for traditional execution: \bdisj{} and \bpushand{}.
%\bdisj{} emulates what real systems might do for predicate expressions with outermost disjunctions.
%The planner treats each child of the corresponding predicate tree's root node as a separate query to be planned.
%Each child is executed independently of the others, applying pushdown optimizations when possible, and a final union operator is appended to combine the results and remove any duplicate tuples.
%\bpushand{} takes a slightly different approach.
%It checks to see if the children share any predicate subexpressions in common.
%If so, those predicate subexpressions are pulled out using a conjunction and only those predicate subexpressions get pushed down.
%The remaining children are executed are all the joins.

%\cref{fig:job-bdisj-combined,fig:job-bpushand-combined,fig:job-bpushand-min,fig:job-overhead} present the results of our evaluation for JOB.
\cref{fig:job}\ifthenelse{\papermode = \cameraready}{ (next page)}{} presents the results of our evaluation for JOB.
We ran each query 5 times for each planner, and the figures use the average of those 5 times.
We present only the total runtimes and do not present planning and executions times separately because in every case, planning time accounted for less than 0.1\% of the total runtime.

\smallskip
\begin{mdframed}
\textbf{Key Takeaways.}
We observed that tagged execution greatly outperformed traditional execution in many cases for JOB.
Specifically, \combined{} had an average 2.7$\times$ speedup over \bdisj{}, and while the results varied more against \bpushand{}, there were still queries in which \combined{} achieved a 19$\times$ speedup over \bpushand{}.
At the same time, we observed that the tagged execution model incurred an average overhead of only 10\% compared to traditional execution.
\end{mdframed}
\smallskip

\ifthenelse{\papermode = \cameraready}{%
  \begin{figure*}
    \centering
    \begin{subfigure}{0.49\linewidth}
      \includegraphics[width=\linewidth]{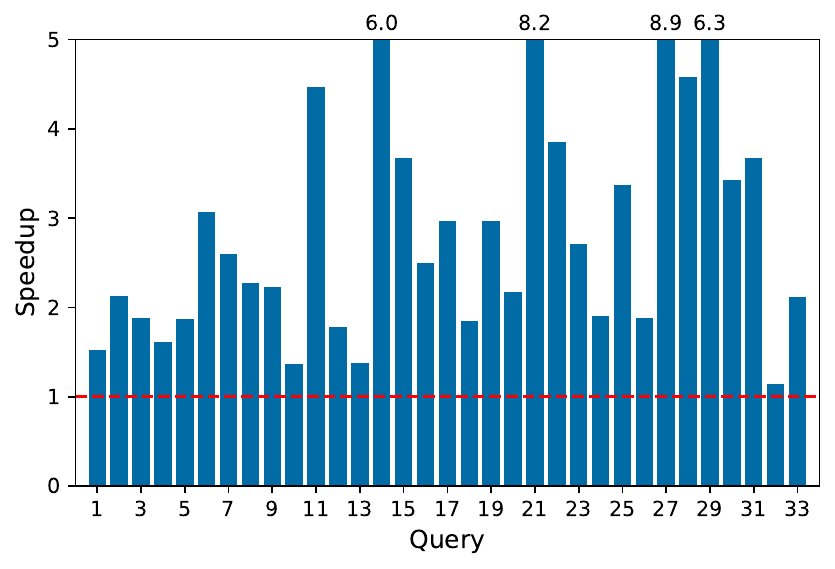}
      \caption{\bdisj{}/\combined{}}
      \label{fig:job-bdisj-combined}
    \end{subfigure}
    \begin{subfigure}{0.49\linewidth}
      \includegraphics[width=\linewidth]{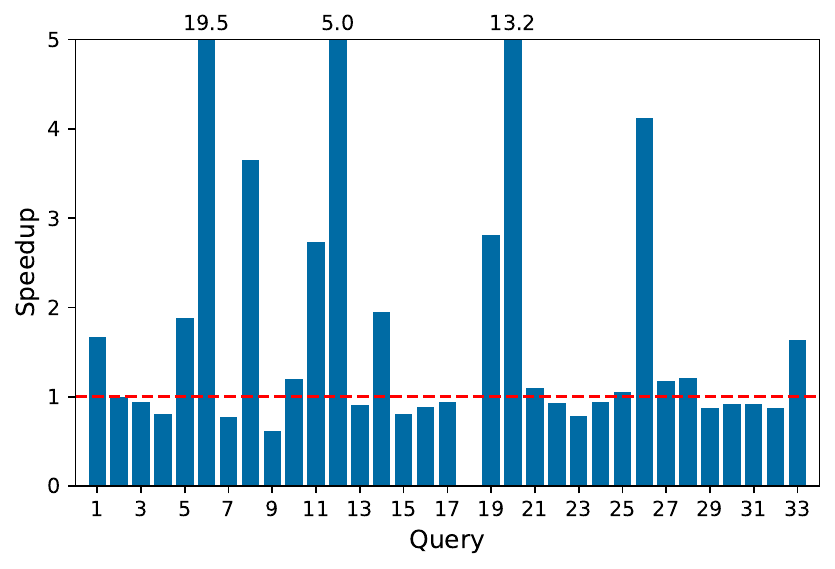}
      \caption{\bpushand{}/\combined{}}
      \label{fig:job-bpushand-combined}
    \end{subfigure}
    \begin{subfigure}{0.49\linewidth}
      \includegraphics[width=\linewidth]{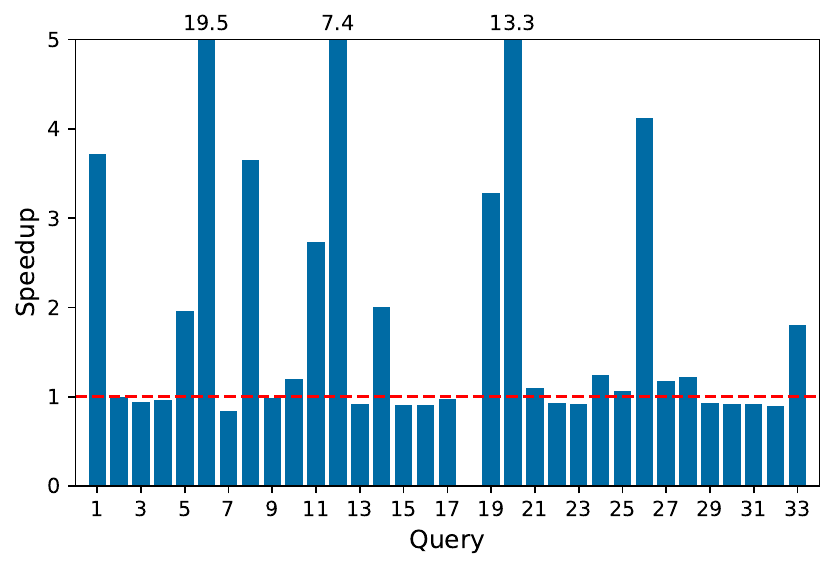}
      \caption{\bpushand{}/\tmin{}}
      \label{fig:job-bpushand-min}
    \end{subfigure}
    \begin{subfigure}{0.49\linewidth}
      \includegraphics[width=\linewidth]{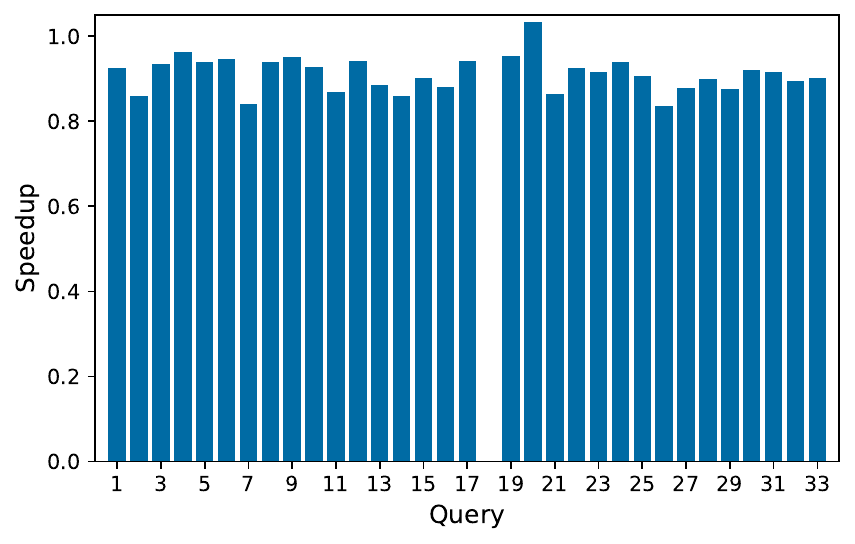}
      \caption{\bpushand{}/\pushand{}}
      \label{fig:job-overhead}
    \end{subfigure}
    \caption{Speedups in total runtime from using tagged execution for JOB queries ($> 1$ means tagged execution is better).}
    %\caption{Potential speedups of total runtime and overhead of using tagged execution for JOB queries.}
    \label{fig:job}
  \end{figure*}
}{}

\cref{fig:job-bdisj-combined} presents the speedups in total runtime achieved by \combined{} over \bdisj{} for each query.
As shown clearly, \combined{} achieved at least a 2$\times$ speedup for most queries and for queries 27 and 21, achieved speedups of 9$\times$ and 8$\times$ respectively.
Deeper analysis revealed a multitude of reasons for these results.
First, the individual queries in each query group of JOB all shared common predicate subexpressions.
However, since \bdisj{} treats \clause[s] as completely independent of one another, the system ended up performing redundant work in evaluating the same predicate subexpression multiple times.
In a similar vein, different \clause[s] often operated on the same set of tuples.
Although \system{} only materializes the indices for each tuple until projection (see \cref{sec:model:impl}), this still meant that the same tuple's indices needed be materialized multiple times in intermediate relations across different \clause[] executions.
The final union operator that \bdisj{} appends to remove duplicate tuples also incurred significant runtime for queries whose final output relation size was large.
In contrast, \combined{} only ever evaluated each predicate subexpression once, and each tuple is only ever materialized once as well.
\combined{} also does not require an additional union operator; the tag system is sufficient to track whether a tuple belongs to the final output or not.
Finally, joins in \combined{} often completed much quicker than the joins in \bdisj{}, even for input (tagged) relations of similar sizes, displaying the importance of selective tag maps.

Since \clause[s] often shared common predicate subexpressions, we wondered what would happen if a traditional execution planner was smart enough to take advantage of that information.
Thus, we searched for common predicate subexpressions that were children to every \clause[] in a query and pulled out those predicate subexpressions to create an equivalent predicate expression with an AND root node (e.g., a predicate expressions $(A \land B \land C) \lor (A \land B \land D)$ would be transformed into $A \land B \land (C \lor D)$).
We ran \bpushand{} and \combined{} on the resulting queries, and \cref{fig:job-bpushand-combined} shows the speedups achieved by \combined{} over \bpushand{}.
Note that \bpushand{} always ran out of memory before completing query 18, so the figure does not include a speedup for that query.
For several queries, \combined{} still achieved significant speedups.
Specifically, for queries 6 and 20, \combined{} achieved speedups of 19$\times$ and 13$\times$ respectively (greater than any speedup achieved over \bdisj{}).
For these queries, \combined{} was able to push down all \patom[s], while \bpushand{} was only able to push down the \patom[s] belonging to the common predicate subexpressions.
This made a large difference because the unpushed \patom[s] included several expensive predicates, such as regular expression matching, and the sizes of the relations after the joins were much larger than the sizes of the base tables.
In addition, the tagged join operator continued to exhibit superior runtimes compared to the regular join operator even for similar sized input (tagged) relations. %, once again displaying the importance of selective tag maps.
On the other hand, for many of the queries, the speedups ranged from 0.9 - 1.1, indicating similar performance between \combined{} and \bpushand{}.
The primary reason for this was due to the highly selective nature of the common predicate subexpressions.
As mentioned, each query group in JOB follows some sort of theme, and the common predicate subexpressions often included highly selective \patom[s] defining that theme.
These highly selective \patom[s] would filter most of the tuples early in the pipeline, preventing \combined{} from really showcasing the benefits of tagged execution.
Even worse, for cases such as query 9, the speedup dipped as low as 0.6$\times$.
However, this was mostly due to inaccurate cost models.
Sometimes, \combined{} would select the ``cheapest'' plan, only for a different tagged execution planner to outperform it.
To account for this, we also measured speedup of the minimum runtime achieved by \emph{any} tagged execution planner over \bpushand{}.
\cref{fig:job-bpushand-min} shows the results.
As can be seen, the minimum speedup becomes 0.8$\times$, and the speedups of several queries jump even higher.
This indicates that with a more realistic cost model, tagged execution could potentially achieve even greater speedups over traditional execution.

Finally, we wanted to measure the ``overhead'' of tagged execution with respect to traditional execution.
That is, given a plan that can be run under both tagged and traditional execution models, how much slower is the tagged execution engine compared to the traditional execution engine in completing that plan.
We accomplished this by looking at the speedups of \pushand{} over \bpushand{}.
The plan generated by these planners forces tagged execution to behave exactly like traditional execution.
Filter operators for all pushed \patom[s] do not include the \negtag{} in their tag maps, and join operators take the full Cartesian product between their input left and right \relslice[s].
\cref{fig:job-overhead} shows the results.
As can be seen, the average speedup is around 0.9$\times$, suggesting a 10\% overhead in using tagged execution over traditional execution.

\subsection{Synthetic Experiments}
\label{sec:eval:synth}

\ifthenelse{\papermode = \submission}{%
\begin{figure*}
  \centering
  \begin{subfigure}{0.24\linewidth}
    \includegraphics[width=\linewidth]{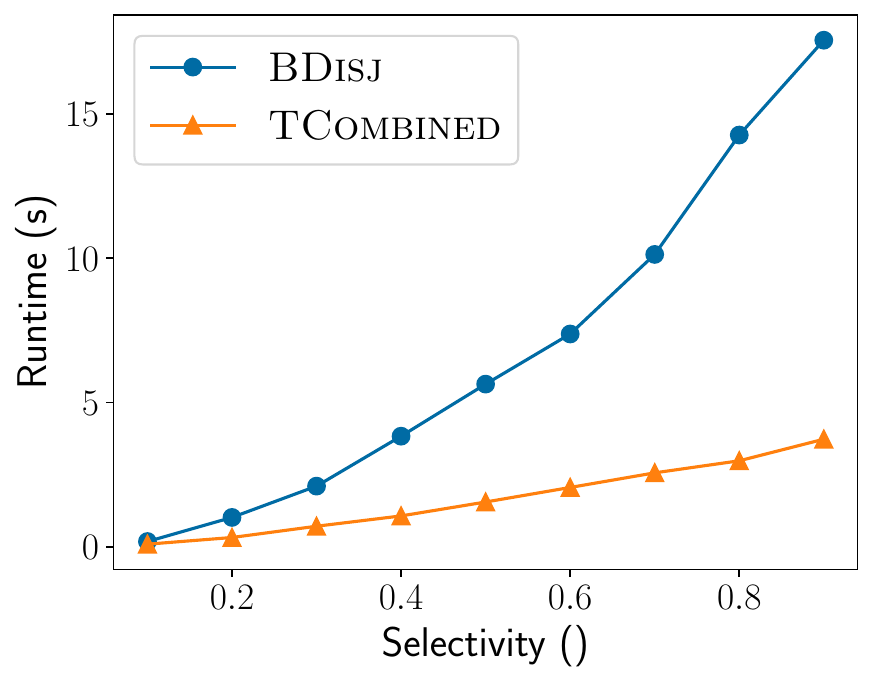}
    \caption{Selectivity (DNF)}
    \label{fig:synth-selectivity}
  \end{subfigure}
  \begin{subfigure}{0.24\linewidth}
    \includegraphics[width=\linewidth]{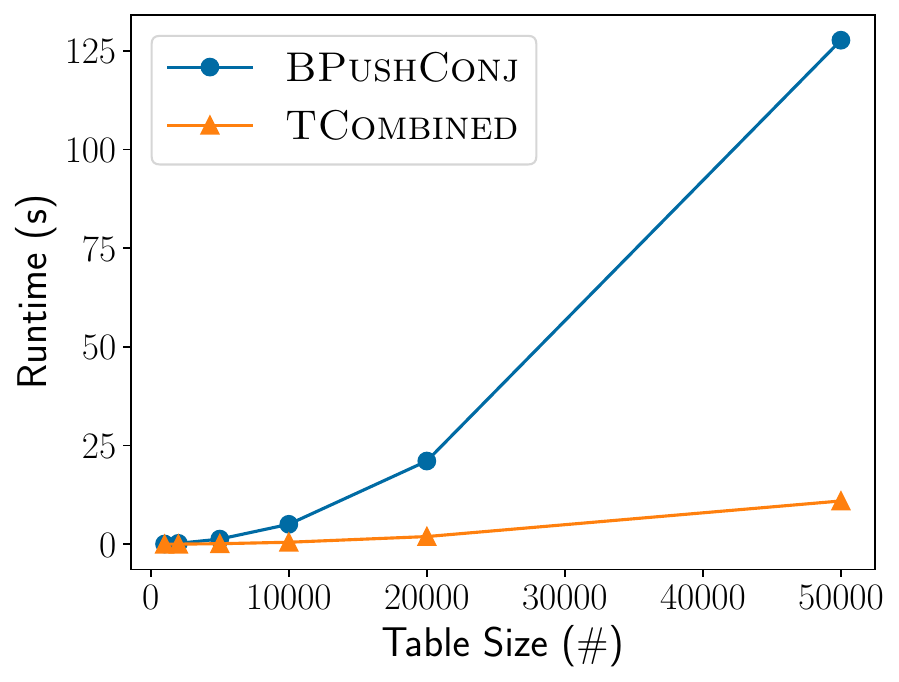}
    \caption{Table Size (CNF)}
    \label{fig:synth-table-size}
  \end{subfigure}
  \begin{subfigure}{0.24\linewidth}
    \includegraphics[width=\linewidth]{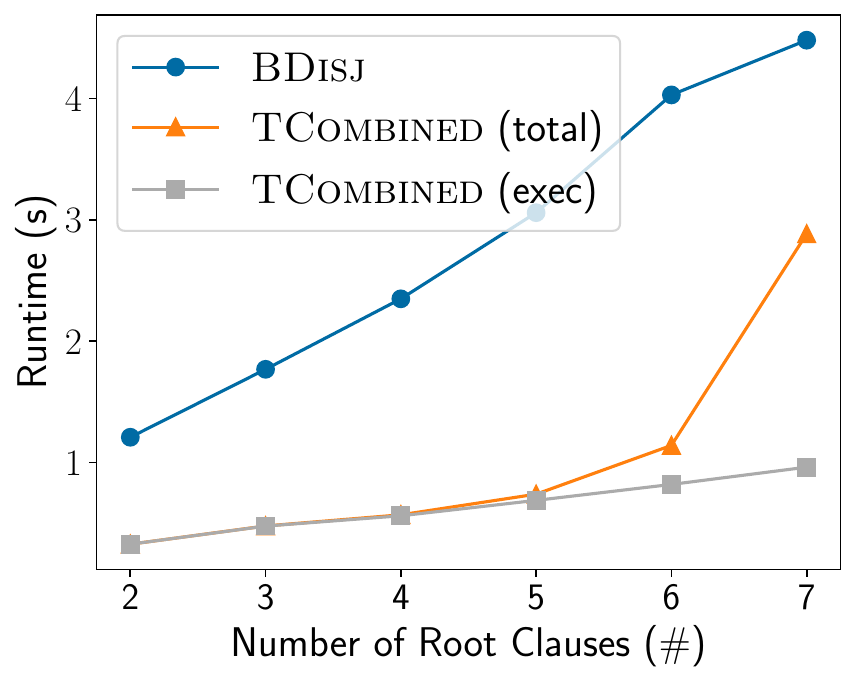}
    \caption{Number of \clause[Cs] (DNF)}
    \label{fig:synth-num-clauses}
  \end{subfigure}
  \begin{subfigure}{0.24\linewidth}
    \includegraphics[width=\linewidth]{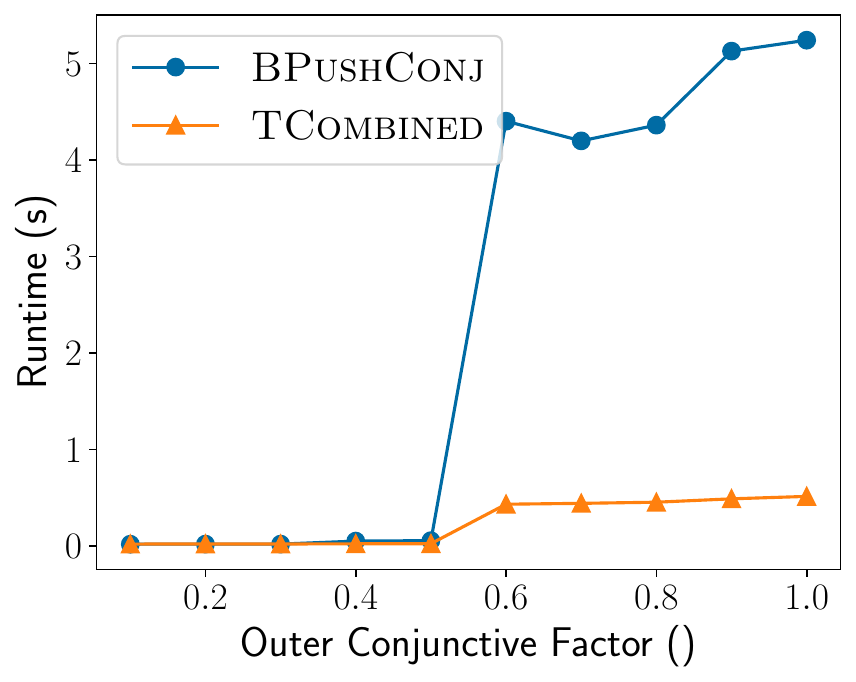}
    \caption{Outer Conjunctive Factor (CNF)}
    \label{fig:synth-outer-conj-factor}
  \end{subfigure}
  \caption{Synthetic experiments varying a number of variety of different parameters.}
  \label{fig:synth}
\end{figure*}
}{%
\begin{figure*}
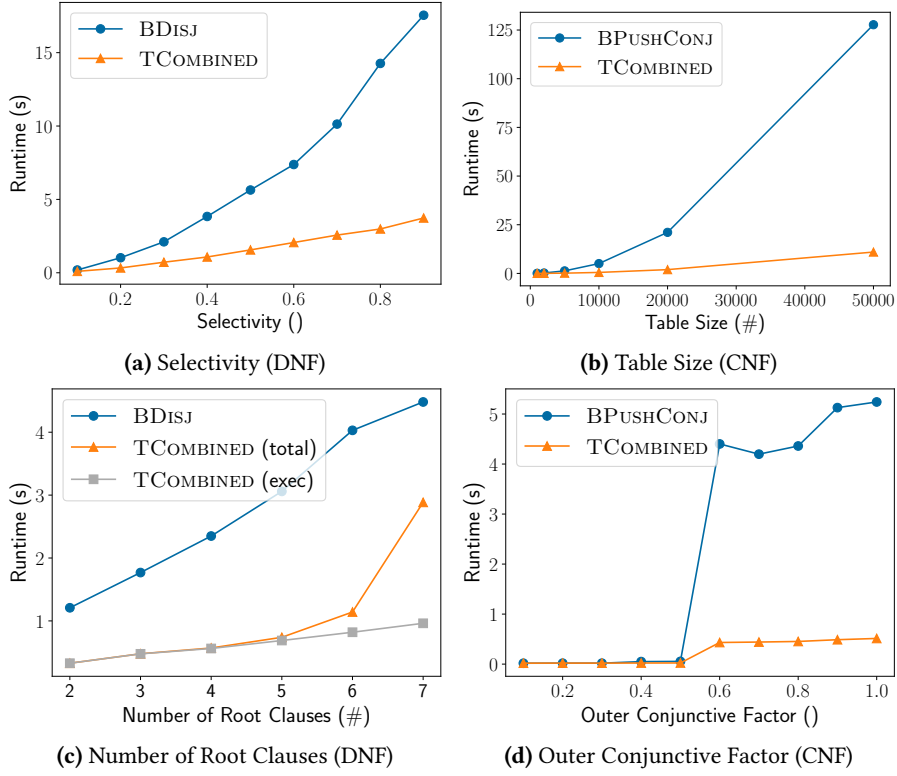

  \centering
  \begin{subfigure}{0.4\linewidth}
    \includegraphics[width=\linewidth]{figs/synth-exp-selectivity-dnf-total_time_ms.pdf}
    \caption{Selectivity (DNF)}
    \label{fig:synth-selectivity}
  \end{subfigure}
  \begin{subfigure}{0.4\linewidth}
    \includegraphics[width=\linewidth]{figs/synth-exp-table_size-cnf-total_time_ms.pdf}
    \caption{Table Size (CNF)}
    \label{fig:synth-table-size}
  \end{subfigure}
  \begin{subfigure}{0.4\linewidth}
    \includegraphics[width=\linewidth]{figs/synth-exp-num_clauses-dnf.pdf}
    \caption{Number of \clause[Cs] (DNF)}
    \label{fig:synth-num-clauses}
  \end{subfigure}
  \begin{subfigure}{0.4\linewidth}
    \includegraphics[width=\linewidth]{figs/synth-exp-outer_conj_factor-cnf-total_time_ms.pdf}
    \caption{Outer Conjunctive Factor (CNF)}
    \label{fig:synth-outer-conj-factor}
  \end{subfigure}
  \caption{Synthetic experiments varying a number of variety of different parameters.}
  \label{fig:synth}
\end{figure*}
}

%SPACE - remove example of CNF

We also evaluated the tagged execution model on a set of synthetic experiments varying different parameters.
These experiments had two base queries; one in CNF and one in DNF.
The DNF version of the query was:
\smallskip
\begin{quoting}[leftmargin=0pt]
  \ttfamily
  \centering
  \begin{varwidth}{1.1\linewidth}
  \begin{tabbing}
  SELECT * FROM \= T0 \= JOIN T1 ON T0.id = T1.fid \\
  \> \> JOIN T2 ON T0.id = T2.fid \\
  \> WHERE\' (T1.A1 < 0.2 AND T2.A1 < 0.2) \\
  \> OR\' (T1.A2 < 0.2 AND T2.A2 < 0.2)
  \end{tabbing}
\end{varwidth}
\end{quoting}
\smallskip
The CNF version of the query was the same, except the ANDs and ORs were swapped in the predicate expressions (i.e., \texttt{(T1.A1 < 0.2 OR T2.A1 < 0.2) AND (T1.A2 < 0.2 OR T2.A2 < 0.2)}).
These are relatively simple queries, but because each \clause[] contains \patom[s] referencing different tables, existing systems cannot optimize them very well.
In fact, there is no way to optimize the CNF version using a traditional execution model.
As for the underlying dataset, the tables \texttt{T0}, \texttt{T1}, and \texttt{T2} each had 10k records.
Table \texttt{T0}'s \texttt{id} attribute was the primary key and had unique values ranging from 1 to 10,000.
\texttt{T1} and \texttt{T2}'s \texttt{fid} attributes were the foreign keys, and their values were randomly generated using a Zipf distribution with a shape parameter value of 1.5.
The attributes that appeared as part of \patom[s] (e.g., \texttt{T1.A1} and \texttt{T2.A2}) had values ranging from 0 to 1, generated uniformly at random.

From these base queries and dataset, we varied a number of various parameters.
For DNF queries, we ran \bdisj{} and \combined{}, while for CNF queries, we ran \bpushand{} and \combined{}.
\cref{fig:synth} shows some of the results.
In every experimental configuration, each query was run 5 times for each planner, and the figures report the average of those 5 times.
With the exception of the experiment varying the number of \clause[s], planning time once again accounted for less than 0.1\% of the total runtime, so only \cref{fig:synth-num-clauses} plots the total runtime and the execution runtime separately.

\smallskip
\begin{mdframed}
\textbf{Key Takeaways.}
%Before delving into the details of each experiment, we first summarize the key takeaways.
We observed that as the number of tuples handled by a query increased, tagged execution increasingly outperformed traditional execution.
Specifically, \combined{} \ifthenelse{\papermode = \techreport}{had}{achieved} speedups of up to 12$\times$ over \bdisj{} for DNF queries and speedups of up to 10$\times$ over \bpushand{} for CNF queries.
%
%For DNF queries, \combined{} achieved speedups of up to 12$\times$ over \bdisj{}.
%For CNF queries, \combined{} achieved speedups of up to 10$\times$ over \bpushand{}.
\end{mdframed}
\smallskip

\textbf{Selectivity.}
The first parameter we varied was the selectivity of the \patom[s], which we varied from 0.1 to 0.9.
\cref{fig:synth-selectivity} presents the results for DNF queries.
As shown, the runtimes diverged drastically as the selectivity value grew, with \combined{} achieving a speedup of 5$\times$ over \bdisj{} when selectivity is 0.9.
When selectivity was small, most tuples were filtered out early, so the difference between \bdisj{} and \combined{} was not as drastic.
However, as the selectivity grew, the sizes of the intermediate and resulting relations also grew, and this had a greater impact on \bdisj{} than \combined{} for three reasons.
%Specifically, as the sizes of the intermediate relations grow, the number of duplicate tuples materialized in \bdisj{} and the amount of work done by the final union operator in \bdisj{} increase, and \combined{} can avoid both of these overheads and improve the performance of join operators using its selective tag maps.
%However, as the selectivity grew, the sizes of the intermediate and resulting relations also grew, and this had a greater impact on \bdisj{} than \combined{} for the same reasons as the JOB experiments.
%
%for three reasons.
First, even though the joins had similar size inputs for both tagged and traditional execution, the selective tag maps in tagged execution reduced the amount of work performed by each join operator.
Second, as selectivity grew, more duplicate tuples were materialized in intermediate relations across different \clause[s] for \bdisj{}, while \combined{} only materialized each tuple once.
Third, the union operator for \bdisj{} handled more tuples with increasing selectivity, resulting in more overhead.
The plot for the CNF version of the query had a slightly different shape.
%As for the result of the CNF version of the queries, the plot has a slightly different shape.
Instead of the runtimes differing as a function of selectivity, there was instead a constant (large) difference between the runtimes for \combined{} and \bpushand{} (\combined{} was faster).
Due to each \clause[] of the CNF referring to multiple tables, \bpushand{} could not push down any \clause[s], so all filter operations were performed after all the joins.
Thus, the time taken to perform the join operations (the largest factor) remained constant even as selectivity varied.

\textbf{Table Size.}
Next, we varied the table sizes of \texttt{T0}, \texttt{T1}, and \texttt{T2} from 1k records up to 50k records.
\cref{fig:synth-table-size} presents the results for CNF queries.
The plot displays the same trends as \cref{fig:synth-selectivity}, with \combined{} achieving a speedup of 12$\times$ when the table size is 50k records.
The reasons were similar as well.
With larger table sizes, the sizes of intermediate and resulting relations also grew, and this affected \bpushand{} more than \combined{}.
Because \bpushand{} could not push any \patom[s] down, it suffered directly from the quadratic growth in the join result.
On the other hand, \combined{} could execute all its filter operators on the base tables, which grew linearly, and the join operators in tagged execution benefited significantly from the use of tag maps.
The plot for the DNF version of the query had the same shape, and the reasons were the same as those mentioned for the DNF version of the selectivity experiment.

%\textbf{Number of Tables.}
%In addition to the table size, we varied the number of tables present in the queries from 3 to 6.
%Each additional table had the same primary key, foreign key relationship to \texttt{T0} as \texttt{T1} and \texttt{T2} (e.g., \texttt{T0.id = T3.fid}).
%However, due to the exponential increase in the size of the join result, we reduced the sizes of all these tables to just 100 records.

\textbf{Number of \clause[Cs].}
We also varied the number of \clause[s] from 2 to 7.
Each additional \clause[] referred to new attributes in \texttt{T1} and \texttt{T2} (e.g., \texttt{T1.A3} and  \texttt{T2.A3}).
\cref{fig:synth-num-clauses} presents the results for DNF queries.
Note that this was the only experiment in which the planning time for \combined{} started to have a noticeable impact on the overall runtime.
As such, \cref{fig:synth-num-clauses} plots \combined{} (total) for the total runtime and \combined{} (exec) for only the execution runtime (without planning).
First, note that the difference in runtime between \bdisj{} and \combined{} (exec) increases with more \clause[s]; at 7 \clause[s], \combined{} achieved a 5$\times$ speedup over \bdisj{}.
Since \bdisj{} executes each \clause[] independently of others, additional \clause[s] meant more tuples materialized multiple times across different \clause[s] and additional join operators, each with their own overhead.
In comparison, \combined{} (exec) only experienced modest increases in runtime from executing the additional \patom[s].
The figure also depicts a significant increase in planning time for \combined{} for more \clause[s].
This increase was due to the unoptimized nature of \pullup{}; recall that \pullup{} attempts to pull up every filter node one node at a time and see if that results in a plan.
As the number of \clause[s] increased, so did the number of filter nodes, and the number of plan comparisons increased exponentially.
%Since \pullup{} initially arranges filter nodes in benefiting order,
A more optimized version of the planner which pulls filter nodes up to the next join juncture could substantially decrease planning time.
The plot for the CNF version of the query also showed an increasing difference in the runtime between the \bpushand{} and \combined{} (exec) (\combined{} was faster), though not as severe as \cref{fig:synth-num-clauses}.
Since \bpushand{} performs all joins before performing any filter operations, even with the base 2 \clause[s], there was a large difference in runtime between \bpushand{} and \combined{}, and increasing the number of \clause[s] only had a secondary effect.

%\textbf{Number of Redundancies.}

\textbf{Outer Conjunctive Factor.}
Throughout our experiments, we observed that if there were selective predicate subexpressions which filtered out most of the tuples early in the pipeline, tagged execution would not have a chance to exhibit its benefits.
In an attempt to isolate and measure this factor, we conducted an experiment in which we varied the selectivity of an additional conjunctive predicate expression, henceforth termed the \emph{outer conjunctive factor}.
For CNF queries, the predicate expression with an outer conjunctive factor of 0.1 had the form: \texttt{T0.A1 < 0.1 AND (T1.A1 < 0.2 OR T2.A1 < 0.2) AND (T1.A2 < 0.2 OR T2.A2 < 0.2)}.
For DNF queries, the same \texttt{T0.A1 < 0.1} was included in each \clause[].
\cref{fig:synth-outer-conj-factor} shows the results for CNF queries.
As we can see, the runtimes for \bpushand{} and \combined{} remain mostly similar until a sharp increase when the outer conjunctive factor is 0.6.
At this point, the tagged execution model finally had a chance to display its benefits, and the difference in runtime peaked when the outer conjunctive factor was 1.0, with \combined{} achieving a 10$\times$ speedup over \bpushand{}.
The reason for the sharp increase at 0.6 was because the first record in \texttt{T0} (i.e., \texttt{T0.id = 1}) had a \texttt{A1} value between 0.5 and 0.6.
As mentioned, we used a Zipf distribution to generate foreign keys randomly, and the most common value that appears in a Zipf distribution is 1.
Thus, the inclusion of the first record when the outer conjunctive factor rose to 0.6 meant a substantial increase in the resulting output and a substantial increase in the runtime for \bpushand{}.
Although the same was true for \combined{}, this inclusion had a much smaller impact on tagged execution due to its efficient join operators with selective tag maps.
The plot for DNF queries had the same shape, for the same reasons.

%\todo{One thing where our implementation differs from the original work is that NULLs are not implemented in our DB, so NULL values are changed to some arbitrary chosen value in our experiments, so this might result in more values being processed than the original dataset}

\todo{We might need a graph/table that measures number of predicates vs number of tags for various queries}

\todo{We probably need some form of memory measurement for the overhead of tagged execution, but we can probably wait for the revision to do that}

%\todo{If I had to add figs, cnf for synth might be good}
%\todo{Perhaps include more figs for tech report}

\section{Related Work}
\label{sec:rel-work}

The set of related work can largely be divided into two parts:
\begin{enumerate*}
  \item works related to tagging
  \item works related to disjunctions
\end{enumerate*}.

\textbf{Tagging.}
Though not explicit, the idea of tagging is present in past works.
Most works in shared work optimization~\cite{chenNiagaraCQScalableContinuous2000}~\cite{giannikisSharedDBKillingOne2012a}~\cite{giannikisSharedWorkloadOptimization2014} \cite{candeaScalablePredictableJoin2009} \cite{krishnamurthyCasePrecisionSharing2004} \cite{hongRulebasedMultiqueryOptimization2009} use some form of tags to keep track of which tuples belong to which queries, and works built on the eddy processor~\cite{avnurEddiesContinuouslyAdaptive2000}~\cite{maddenContinuouslyAdaptiveContinuous2002}~\cite{hellersteinAdaptiveQueryProcessing2000} typically use a tag to keep track of which operators have been evaluated for each tuple.
However, the context in which tags are used for these works is clearly very different compared to our work.
In addition, another critical difference between these works and tagged execution is the \emph{semantic} information present in the tags for tagged execution.
The tags used in these works all represent some sort of simple membership into a query/operator set, and it is difficult to see that as semantic information compared to the predicate results present in the tags for tagged execution.

%However, a critical difference between these works and tagged execution is the \emph{semantic} information present in the tags for tagged execution.
%The tags used in these works all represent some sort of simple membership into a query/operator set, and it is difficult to see that as semantic information.
%Rather, the predicate results present in the tags for this work and the layout information discussed in \cref{sec:disc} serve as better examples.
%Another crucial difference is that almost all of these works make no mention of disjunctions.
%Only the CACQ work by Madden et al.~\cite{maddenContinuouslyAdaptiveContinuous2002} makes a brief mention of disjunctions but offers no real details.

\textbf{Disjunctions (Bypass).}
The most relevant related work is the line of work regarding the bypass technique~\cite{kemperOptimizingDisjunctiveQueries1994}~\cite{steinbrunnBypassingJoinsDisjunctive1995}~\cite{claussenOptimizationEvaluationDisjunctive2000}.
As originally introduced by Kemper et al.~\cite{kemperOptimizingDisjunctiveQueries1994}, the bypass technique augments filter operators in traditional execution with an optional ``false'' output stream.
Input tuples which evaluate to true are output to the regular ``true'' output stream, and those that evaluate to false are sent to the ``false'' output stream.
The desire is that the tuples in the ``true'' output stream bypass the more expensive filter operators that appear later in the pipeline.
Steinbrunn et al.~\cite{steinbrunnBypassingJoinsDisjunctive1995} extend this idea to join conditions, and Claussen et al.~\cite{claussenOptimizationEvaluationDisjunctive2000} expand the technique to include NULL values.
While similar, tagged execution different from bypass in two main ways:
(1) First is the reuse of query operators and shared work.
By using tags to encode predicate expression results, tagged execution can reuse the same query operators for different \relslice[s] with different predicate expression results.
%As \cref{fig:query-plan} shows, tagged execution only scans each table once and uses a single join operator to execute the query.
However, bypass depends on traditional relational query operators, and this results in using different query operators for relations with different predicate expression results.
Thus, even for a query as simple as \cref{query:ex}, bypass requires multiple scans of the input tables and multiple join operators to execute.
(2) Second is the separation of tag space and query plan space.
Tagged execution encapsulates the evaluation of the predicate predicate in tags and separates this tag space from the query plan.
%As a result, tagged execution planners can make planning decisions, such as choosing to push down/pull up filters, independently of the predicate expression evaluation.
As a result, tagged execution planners can make planning decisions independently of the predicate expression evaluation.
On the other hand, bypass embeds the predicate expression evaluation directly into the query plan and only produces plans in which predicates are all pushed down.
However, as discussed in \cref{sec:plan,sec:eval}, this is not always desirable.

\textbf{Disjunctions (CNF/DNF).}
Aside from the bypass technique, most works involving disjunctive predicate expressions typically focus on converting the predicate expression into either CNF or DNF and optimizing the execution from these forms~\cite{selingerAccessPathSelection1979}~\cite{straubeQueriesQueryProcessing1990}~\cite{jarkematthiasQueryOptimizationDatabase1984}~\cite{muralikrishnaOptimizationMultiplerelationMultipledisjunct1988}~\cite{changOptimizationDisjunctiveQueries1997}.
\bdisj{} and \bpushand{} in our evaluation serve as representatives for these works.
However, in addition to the inefficiencies of CNF/DNF-based methods highlighted in the introduction, it is well-known that conversion into CNF/DNF can result in an exponential number of terms~\cite{russellArtificialIntelligenceModern2016}, so just transforming the input into the correct form can be quite expensive.
In comparison, tagged execution has no such requirements on the form of the input predicate expression and optimizes all predicate expression forms equally well.

\textbf{Disjunctions (Factorization).}
Chaudhuri et al.~\cite{chaudhuriFactorizingComplexPredicates2003} introduce techniques to factorize a disjunctive predicate expression to maximize usage of existing indexes.
In a sense, factorization works in the ``opposite'' direction of tag generalization, and we could have used factorization-like techniques to reduce the tag space in our work.
However, the techniques presented by Chaudhuri et al. involve searching over an exponential space, as opposed to \propup{}'s linear runtime, and tag generalization has the highly convenient benefit of exposing which tags can be discarded early.

\textbf{Disjunctions (Ordering).}
The final group of works on disjunctions all have to do with ordering \patom[s]~\cite{hananiOptimalEvaluationBoolean1977}~\cite{kastratiOptimizationDisjunctivePredicates2017}~\cite{kastratiGeneratingOptimalPlans2018}~\cite{kimOptimizingQueryPredicates2022}.
Given a query with a disjunctive predicate expression, these works attempt to find the best order to evaluate the \patom[s] of that predicate expression.
However, these works all assume a single-table query with no joins, so their relevance to our work is limited.

%\textbf{Query Rewriting.}
%Our work shares some similarities with query rewriting.
%However,

\section{Conclusion}
\label{sec:conc}

In this paper, we presented the tagged execution model, a powerful new way to execute queries, and demonstrated how it can be used to optimize disjunctive queries.
We showed how we could reduce the tag space using tag generalization and introduced several new planners which can take advantage of the benefits offered by tagged execution.
Finally, in our evaluation, we showed that our tagged execution planners outperform traditional execution planners with an average speedup of 2.7$\times$ and a maximum speedup of 19$\times$ in certain situations, while only incurring an average overhead of 10\%, highlighting the prowess of the tagged execution model.

\ifthenelse{\papermode = \techreport}{%
  \pagebreak
  \appendix

\section{Benefit Score}
\label{sec:benefit}
The benefit score estimates the value of applying a certain filter operator before others.
It differs from other estimates of filter operator importance, such as simple selectivity and Boolean Difference Calculus~\cite{kemperOptimizingBooleanExpressions1992}, in that it is calculated with respect to a set of unapplied filter operators.
%This allows for more accurate estimates
This is important for cases in which we want an ordering for a \emph{subset} of all filter operators.
In such cases, we only care about the impact each filter operator has on other filter operators in the subset and not on all other filter operators.
By specifying a reference unapplied filter set, the benefit score gives more accurate estimates of the value of each filter operator in such situations.
%To give an example, in \cref{query:ex}, after pushing all filter operators down to the base table level, we must determine an ordering for the predicates \texttt{t.year > 1980} and \texttt{t.year > 2000} in the \texttt{title} table.
%In this case, when estimating the value of applying \texttt{t.year > 2000} before \texttt{t.year > 1980}, the of applying \texttt{t.year > 2000} before \texttt{mi\_idx.score > '7.0'} is irrelevant, since
The algorithm to calculate the score is given by \cref{alg:benefit}\footnote{The exact algorithm additionally allows specifying the number of input tuples for each filter operator in the unapplied reference filter operator set. This effectively weights each filter operator with a different level of importance. However, we omit such details here.}.

\begin{algorithm}
  \begin{algorithmic}[1]
    \Require Filter operator to score ``to\_score'', Unapplied filter operator set ``unapplied\_set''
    \State benefit $\gets$ 0
    \For{unapplied $\in$ unapplied\_set}
    \State is\_and\_descendant $\gets$ true
    \State is\_or\_descendant $\gets$ true
    \For{ancestor\_path $\in$ ancestor\_paths(unapplied)}
    \If{$\forall \text{parent} \in \text{parents(to\_score)}, \text{parent} \not\in \text{ancestor\_path} \lor \text{isOr(parent)}$}
    \State is\_and\_descendant $\gets$ false
    \EndIf
    \If{$\forall \text{parent} \in \text{parents(to\_score)}, \text{parent} \not\in \text{ancestor\_path} \lor \text{isAnd(parent)}$}
    \State is\_or\_descendant $\gets$ false
    \EndIf
    \EndFor
    \If{is\_and\_descendant}
    \State benefit $\gets$ benefit + (1 - selectivity(to\_score))
    \EndIf
    \If{is\_or\_descendant}
    \State benefit $\gets$ benefit + selectivity(to\_score)
    \EndIf
    \EndFor
    \State \Return benefit
  \end{algorithmic}
  \caption{\calcbenefit{}}
  \label{alg:benefit}
\end{algorithm}

For each filter operator in the unapplied set, the algorithm searches to see if that filter operator exists as a descendant of to\_score's parent in the predicate tree.
If so, then the benefit score is updated with (1 - selectivity(to\_score)) if to\_score's parent is an AND node or just selectivity(to\_score) if to\_score's parent is an OR node.
This is because if to\_score's parent is an AND node and we apply to\_score first, then we only need to consider the tuples which evaluated to true for to\_score as as the input to unapplied\footnote{This assumes that the predicates $P$ and $\neg P$ are treated as different predicate subexpressions in the predicate tree.}.
Thus, we can remove (1 - selectivity(to\_score)) of the tuples from consideration by applying to\_score first, and that is the ``AND benefit'' of to\_score with respect to this unapplied filter operator.
Similarly if to\_score's parent is an OR node and we apply to\_score first,  then we only need to consider the tuples which evaluated to false for to\_score as as the input to unapplied.
Thus, we can remove selectivity(to\_score) of the tuples from consideration by applying to\_score first, and that is the ``OR benefit'' of to\_score with respect to this unapplied filter operator.

However, due to the fact that a predicate subexpression may appear multiple times in the predicate tree, \cref{alg:benefit} appears much more complex.
Here, to\_score may have multiple parents (one for each instance in the predicate tree), and one instance of unapplied may be a descendant while another is not.
Thus, \cref{alg:benefit} shows the modifications necessary to handle these cases.
The function ancestor\_paths(unapplied) retrieves the path to the root node for each instance of unapplied in the predicate tree, and parents(to\_score) returns the parent of each instance of to\_score in the predicate tree.
If one of to\_score's parents appears in every ancestor path of unapplied, then it is possible that to\_score has some effect on unapplied.
Specifically, if an AND parent of to\_score appears in every ancestor path of unapplied, then unapplied receives the ``AND benefit'' of applying to\_score first.
Similarly, if an OR parent of to\_score appears in every ancestor path of unapplied, then unapplied receives the ``OR benefit'' of applying to\_score first.
Note that it is possible to receive both the ``AND benefit'' and ``OR benefit'' of applying to\_score first; in this case, unapplied does not have to be applied at all.
The algorithm calculates this benefit value for each filter operator in the unapplied set and returns the sum.

Sorting a set of filter operators according to ``benefiting order'' calculates the benefit score of each filter operator with respect to the remaining filter operators and sorts those filter operators in decreasing $\left< benefit \right> / \left< cost\mhyphen factor \right>$ order.

}{}

\ifthenelse{\papermode = \techreport}{%
  \bibliographystyle{plainnat}
  \pagebreak
}{%
  \bibliographystyle{ACM-Reference-Format}
}
\bibliography{main}

\end{document}